# Deterrence Effects of Social Media Interventions on Health Misinformation Dissemination by Bots and Humans

Amir Karami[a*], Nima Kordzadeh[b], Serena Harn[c]

[a*]School of Data Science and Analytics, Kennesaw State University, akarami@kennesaw.edu
[b]The Business School, Worcester Polytechnic Institute
[c]School of Information Science, University of South Carolina

**Abstract:** In the realm of social media, information dissemination is pivotal, yet it is tainted by the proliferation of misinformation propagated by both bots and humans, bearing consequential impacts on individuals and society. To address this issue, social media platforms have implemented removal, reduction, and informing interventions, acting as deterrent mechanisms to dissuade users from engaging in the spread of misinformation. Nonetheless, the sustained effectiveness of these interventions on bots and humans remains unclear. Drawing on deterrence theory, this study examines the efficacy of social media interventions on bots and humans sharing health misinformation. Our results show that most interventions can have sustained effects on bots and their activities for years after intervention implementation. However, the interventions may not have significant deterrence effects on humans and their activities. Our findings offer important theoretical and practical implications by highlighting the importance of studying both bots and humans and developing creative strategies to tackle health misinformation dissemination.

## 1- Introduction

The repercussions of health misinformation through social media are significant and far-reaching. Such misinformation, consisting of false, inaccurate, or misleading content (L. Wu et al., 2019), can spread rapidly across platforms, *where individuals increasingly depend on user-generated content that may lack credibility when making health-related decisions* (Y. Wang et al., 2019). This can lead to unnecessary or inappropriate healthcare utilization, contributing to a $9 billion burden on the U.S. healthcare sector (CHEQ, 2019). Moreover, health misinformation perpetuates myths and misconceptions, even after they have been debunked, thereby undermining public trust in the healthcare system (The Lancet Infectious Diseases, 2020), straining patient–physician relationships (Bauchner, 2019), impeding evidence-based medicine (Hill et al., 2019), and in extreme cases contributing to mortality risks (S. B. Johnson, Park, Gross, & James, 2018). A glaring example is the contribution of misinformation campaigns promoting anti-vaccine beliefs to the 2019 measles outbreaks in the U.S. (N. F. Johnson et al., 2020). Thus, it is important to combat health misinformation to protect the health and well-being of individuals and society, as well as preserve trust in institutions and promote informed decision-making.

Misinformation on social media can be propagated by both bot and human users. Bots, which are automated programs designed to post content and interact with users, can significantly amplify the spread and perceived legitimacy of misinformation, complicating efforts to manage its dissemination (Salge et al., 2022). These bots are capable of mimicking human behaviors, influencing users' exposure to and sharing of misleading content (Shao et al., 2017). Humans, on the other hand, tend to cluster around shared beliefs, exchanging and reinforcing misinformation within closed networks. This dynamic contributes to polarization and the persistence of misinformation, as dissenting views remain underrepresented or absent (Cinelli et al., 2021). Therefore, it is crucial to prevent both bots and humans from disseminating misinformation.

To combat misinformation, social media platforms have implemented removal, reduction, and informing interventions, such as removing content, suspending user accounts, limiting the visibility of malicious activities, and attaching warning labels to posts (Ng et al., 2021; Zannettou, 2021; Karami, 2025). These interventions aim to reduce the spread of false information and mitigate potential harm in real-world scenarios, such as public health crises (Loomba et al., 2021). However, the comprehensive impacts of these interventions are not yet fully understood, particularly with respect to their sustained effects on both bots and humans. This limitation brings to light two related research gaps that require further investigation.

First, there is a noticeable lack of studies that differentiate the impacts of interventions on bots versus humans. Distinguishing between these groups is essential because they play fundamentally different roles in the information ecosystem. Bots can amplify misinformation at scale, distort the perceived popularity of false content, adapt rapidly to platform interventions, and manipulate algorithmic visibility (Shao et al., 2017; Y. Zhang, Song, Koura, et al., 2023). Human users, by contrast, usually spread misinformation through slower, socially embedded processes such as repeated sharing within communities (Bak-Coleman et al., 2022; Fried, 2019; Siebert & Siebert, 2023). If studies do not distinguish between these groups, findings may be skewed, especially if one group dominates misinformation activity. While the goal is to help humans avoid false information and prevent bots from spreading it, most research (Himelein-Wachowiak et al., 2021; Shao et al., 2017) focuses only on overall misinformation volume, without accounting for these critical differences.

Second, there is a significant oversight in assessing the sustained effects of these interventions, as most research overlooks the circulation of misleading content over time. Although many studies document short-term changes following intervention rollouts, far less is known about whether these effects persist or fade as misinformation actors adjust their behaviors (Shao et al., 2018). Moreover, short-term declines can be misleading, as misinformation may rebound once actors adapt to the intervention and develop ways to circumvent it, underscoring the need to evaluate whether intervention effects endure beyond their initial impact (Sharma et al., 2024). A lack of long-term efficacy can ultimately undermine confidence in platforms and the interventions they implement. As such, understanding these patterns and their long-term effects is critical.

Without evaluating both the differential behaviors of bots and humans and the sustained effects of interventions, platform strategies may appear effective in aggregate while masking important user-type differences or producing only short-lived outcomes. To address these gaps, this study applies deterrence theory, viewing platform interventions as deterrent mechanisms, to examine how removal, reduction, and informing strategies influence bots and humans in the context of health misinformation. The two research questions guiding this study are as follows:

1- To what extent do removal, reduction, and informing interventions exert sustained deterrence effects on the number of bots and their activities related to health misinformation on social media?

2- To what extent do removal, reduction, and informing interventions exert sustained deterrence effects on the number of human accounts and their activities related to health misinformation on social media?

This study centers on Twitter[1] interventions targeting health misinformation due to its sustained presence and serious consequences for public health. Unlike political or disaster-related misinformation, which often spikes around specific events and then declines, health misinformation tends to persist, with certain topics (e.g., unproven cancer treatments) resurfacing repeatedly over time (Muhammed T & Mathew, 2022). This persistence makes it particularly well-suited for examining the long-term effects of intervention strategies. Misleading health claims can lead to harm, such as forgoing effective treatment in favor of unverified alternatives (Kisa & Kisa, 2025). These risks are especially pronounced among marginalized populations, where health misinformation compounds existing disparities and deepens structural inequities (Ford et al., 2013). Beyond individual health decisions, health misinformation also undermines trust in healthcare institutions, making people more hesitant to engage with reliable medical guidance. Because of this sustained impact and measurable harm, professional fact-checkers often give health-related misinformation priority in their assessments (Sehat et al., 2024).

This study makes the following contributions to the information systems literature on misinformation interventions on social media platforms. First, the study applies deterrence theory to conceptualize the mechanisms through which removal, reduction, and informing strategies influence user behavior, contributing to the theoretical development of intervention efficacy in digital environments. Second, it offers a longitudinal, theory-driven analysis of platform interventions over a decade. The analysis addresses the temporal limitations of prior studies by examining the sustained effects of the three interventions. Third, it introduces a user-type specific perspective by distinguishing between bots and human actors, an important yet underexplored dimension of social media dynamics. From a practical standpoint, the findings provide crucial insights for platform providers, policymakers, and regulators by evaluating how these interventions influence users over time.

## 2- Theoretical background

### *2.1 Deterrence theory and social media interventions*

Deterrence theory originated in classical criminology theory and has offered a foundation for criminal justice policies and practices (Tomlinson, 2016). This theory posits that punishment can effectively deter individuals from engaging in undesirable behavior (Beccaria, 2016). Closely connected to the core tenet of rational choice theory, this theory assumes that people are rational decision-makers who weigh the potential costs and benefits of their actions (Williams & Hawkins, 1986). Deterrence theory argues that if punishment is severe, certain, and swift, the presence of a perceived risk of punishment can discourage individuals from behaving in a particular manner (Abramovaite et al., 2023). This perception is based on the belief that the disadvantages and potential perils associated with engaging in undesirable behavior outweigh any potential gains. As a result, potential aggressors are deterred from taking action due to the anticipated negative consequences.

[1] Twitter underwent a name change to X in 2023. Since our data was collected prior to this name change, we refer to it as Twitter in this paper.

Deterrence can be conceptually separated into two distinct effects: specific (direct) deterrence and general (indirect) deterrence (Stafford & Warr, 1993). Specific deterrence aims to dissuade an individual or entity, who has previously exhibited a specific behavior, from repeating it in the future. The objective is to escalate the costs associated with repeating the behavior to a level where the individual or entity is discouraged from doing so. For example, when an individual is convicted of driving while intoxicated (DWI), specific deterrence measures are implemented to discourage them from repeating the offense. These measures can include penalties such as license suspension. The aim is to deter the individual and others from engaging in DWI, promoting compliance with the law and ensuring road safety (Lee & Teske Jr, 2015). General deterrence strives to dissuade prospective offenders from participating in undesirable actions by ensuring the potential consequences of such behavior are widely recognized. This goal is often achieved by enforcing conspicuous and harsh penalties for particular offenses, acting as a deterrent to potential individuals who might contemplate engaging in similar actions. For example, the implementation of new countermeasures targeting driving under the influence of alcohol establishes a widespread perception of potential punishment. The general threat of sanctions acts as a deterrent, discouraging individuals from engaging in drunk driving and resulting in a reduction of injuries and fatalities in collisions related to alcohol consumption (Byrne et al., 2016). Deterrence theory has also been adopted to explain behavior in various information systems contexts, including cyberloafing (Hensel & Kacprzak, 2021), hospital information exchange (Gefen et al., 2019), cybersecurity (D'arcy & Herath, 2011), software piracy (Higgins et al., 2005), and offensive behavior on social media (Seering et al., 2017).

The interventions aimed at addressing misinformation on social media platforms can be analyzed through the lens of deterrence theory. Social media platforms typically employ two levels of interventions: (1) content-based and (2) user-based (Karami, 2025). The first level revolves around individual pieces of content, involving actions like labeling misleading content, displaying warnings to individuals prior to sharing or reacting to the content, reducing the visibility of the content and/or preventing it from being recommended, providing links for additional information concerning misleading content, disabling interaction features, and deleting content containing misinformation (Twitter, 2023). The second level focuses on educating or limiting users to control the spread of misinformation. This entails highlighting misinformation, featuring informative messages to counter misleading narratives, and locking (suspending) accounts (Karami, 2025). While bot accounts share (mis)information on social media platforms automatically, they are also developed by individuals with computer skills to simulate human behaviors (Boshmaf et al., 2013). The creators of social bots adapt their methods and strategies for different reasons, such as evading detection (Hayawi et al., 2023). Thus, the interventions implemented on social media can also act as a deterrent to bot creators.

The specific deterrence effect of interventions, both at the user and activity levels, directly influences users (Stafford & Warr, 1993), including bot creators, who have been informed or punished as a deterrent against further dissemination of misinformation. This effect entails punitive consequences (e.g., suspending) or learning opportunities (e.g., labeling) that discourage them from repeating such behavior (Karami, 2025). The general deterrence effect operates indirectly on involved users who share misinformation and are or are not punished or informed, as well as uninvolved users who have the potential to spread misleading content. The general deterrence effect functions in two ways. First, it establishes an environment wherein involved and uninvolved users are discouraged from engaging in sharing misinformation due to the perceived risks and consequences associated with their actions. As intervention strategies (e.g., account suspension) are visible and are discussed on news outlets, social media actors are provided with evidence of (in)appropriate behavior (Seering et al., 2017). Second, this can be explained by the validity effect. According to the validity effect, multiple exposures to misinformation can bolster belief in its validity (Boehm, 1994). Intervention efforts (e.g., suspending accounts or removing false content) disrupt this cycle, reducing the chance for repeated exposure, and thereby diminishing the influence of misinformation, safeguarding uninvolved users, and fostering a more reliable information ecosystem.

Interventions can have different combinations of immediate and sustained deterrence effects. The immediate deterrence effect pertains to short-term outcomes observed shortly after implementing social media interventions (Jhaver et al., 2021). It focuses on the initial effectiveness and behavioral changes resulting from the intervention during its early stages. In contrast, the sustained deterrence effect considers the long-term impact of these interventions on users and their activities (Schaffer et al., 2021). It assesses whether the initial changes in behavior induced by the intervention are maintained over time. It is worth noting that not all interventions necessarily have both immediate and sustained effects, as some may have an immediate impact without a sustained one (Hanbury et al., 2013). Therefore, a nuanced understanding of these effects is essential for informed decision-making in the development and implementation of interventions (Clarke et al., 2019). This study examines sustained deterrence effects (Schaffer et al., 2021), contributing to our understanding of the lasting effectiveness of deterrence interventions and their potential to lead to sustained changes in user behavior or a long-term reduction in targeted misinformation sharing.

### *2.2 Health misinformation: A growing concern*

Misinformation on social media is pervasive and can be categorized into three broad domains: disaster, politics, and health (Muhammed T & Mathew, 2022). Each of these domains operates differently in terms of lifespan, emotional appeal, and public impact. Disaster misinformation tends to be event-driven and short-lived, often peaking during crises such as natural disasters or pandemics and then quickly fading as the event passes (Zhai et al., 2024). Political misinformation, while it may persist over election cycles or ideological debates, is often polarized and tied to news cycles, policy debates, or leadership changes (Angelucci et al., 2024). In contrast, health misinformation is both persistent and emotionally charged, making it particularly insidious and difficult to manage. It often resurfaces over long periods, such as recurring myths about cancer cures, and therefore lends itself well to longitudinal study (Muhammed T & Mathew, 2022).

Health misinformation is characterized by two key elements: dread and wish (Chua & Banerjee, 2018). Dread-based misinformation often instills fear by suggesting that certain behaviors, such as consuming specific foods, can lead to severe health risks. In contrast, wish-based misinformation offers hope by claiming that these same behaviors can miraculously treat diseases. These emotional levers make health misinformation especially compelling and difficult to counter. Moreover, its consequences are often immediate and tangible; misinformed decisions about treatment can result in direct harm, unlike political misinformation, which generally influences belief systems or voting behavior, or disaster misinformation, which may misguide short-term safety decisions (Muhammed T & Mathew, 2022). Given these characteristics, health misinformation warrants focused attention in both academic research and platform-level intervention strategies (Kisa & Kisa, 2025).

Health misinformation is disseminated by both human users and automated bot accounts (Broniatowski et al., 2018; Dunn et al., 2020). Bot prevalence can range from 0.86% (Egli et al., 2021) to 45% (Young, 2020). Bots interact not only with other bots but also with humans to promote narratives, individuals, or organizations (Broniatowski et al., 2018; Persily, 2017). They often attempt to present themselves as influential users to better engage with humans (Y. Zhang et al., 2023). This dual engagement with both bots and humans allows them to effectively disseminate content and influence social media dynamics (Gilani et al., 2017). Distinguishing bot accounts from human accounts is challenging because bots can closely mimic human behavior (Hagen et al., 2022; Y. Zhang et al., 2023) and people overestimate their ability to identify false information (B. A. Lyons et al., 2021). As a result, humans often engage with not only humans but also bots on social media (Gilani et al., 2017). This is particularly true for bots that promote content aligned with users' interests, making them more likely to attract human interaction (Wischnewski et al., 2024). In short, there are engagements within bot networks, within human networks, and across bot-human interactions, all of which contribute to the spread and persistence of health misinformation.

The impact of misinformation in the context of healthcare, whether shared by bots or human users, is profound and concerning (Laato et al., 2020). False information can lead individuals to make critical decisions about their health that lack a factual basis. For example, a person diagnosed with a life-threatening condition, such as cancer, may experience anxiety and turn to social media for information and support. They may be directed to articles or websites offering hope and promoting alternative treatments, which may not be scientifically substantiated for such conditions. This misinformation can lead to individuals opting for unconventional treatments over established ones, potentially resulting in a mortality rate 2.5 times higher than those who choose conventional therapies like chemotherapy (S. B. Johnson, Park, Gross, & Yu, 2018). Despite sounding too good to be true, such information may gain credence through testimonials from patients or their loved ones claiming remarkable recoveries. Paradoxically, a survey conducted by the American Society of Clinical Oncology revealed that approximately 40% of Americans believe in the efficacy of alternative therapies for curing cancer (Cavallo, 2019). This further underscores the urgency of addressing health misinformation in the digital age.

Health misinformation is not limited to highly viral topics (Broniatowski et al., 2022) and encompasses a wide range of topics, including COVID-19, general health, cardiovascular disease, diet and nutrition, smoking, water safety, drugs, cancer (Y. Wang et al., 2019), pregnancy (Bryant et al., 2014), vaccines (Z. Xu & Guo, 2018), and influenza (B. Chen et al., 2018). For example, studies focusing on cancer misinformation have investigated misinformation about cancer in general (DiFonzo et al., 2012; S. B. Johnson et al., 2022), breast cancer (Wilner & Holton, 2020), gynecologic cancer (L. Chen et al., 2018), prostate cancer (Loeb et al., 2019; A. J. Xu et al., 2021), cancer screening (Okuhara et al., 2017), and cancer nutrition (Warner et al., 2022). The prevalence of cancer misinformation on social media was estimated from 11% (Wilner & Holton, 2020) to 98% (A. J. Xu et al., 2021), with misinformation often receiving more engagement than true or high-quality information (S. B. Johnson et al., 2022; Loeb et al., 2019).

### *2.3 Social media intervention and health misinformation*

In general, social media platforms use three types of interventions to tackle misinformation: removal, reduction, and informing (Center for an Informed Public et al., 2021). Starting in 2017, Twitter launched the removal strategy to combat the threat posed by malicious actors, including examining reported users or content and managing linked accounts that violated Twitter's rules (Crowell, 2017; Harvey & Gasca, 2018; Roth & Harvey, 2018; Twitter, 2017). Malicious actors include both automated bots, such as traditional spambots, social spambots, content polluters, and fake followers, and human users, including trolls who misrepresent their identities to foster discord and spread misinformation on topics ranging from politics to health issues like cancer (Broniatowski et al., 2018; Jamison et al., 2019; Karami et al., 2021). Twitter implemented the removal strategy to target active misinformation spreaders, including high-profile influencers and those posting low-quality tweets infrequently, as well as followers of suspicious accounts (Jhaver et al., 2021; Roth & Harvey, 2018).

In 2018, Twitter added the reduction strategy, which involved diminishing the visibility of suspicious accounts in follower lists, changing account statuses to read-only to limit engagement, revising account metrics of suspicious or malicious actors (e.g., adjusting follower counts), banning political advertisements, and restricting the reach of misleading content (Roth & Harvey, 2018; Roth & Pickles, 2020). In 2021, Twitter also developed its informing strategy, which included sponsoring and hosting media literacy campaigns, labeling misleading content, such as those related to the 2020 election and the COVID-19 pandemic, and allowing users to contribute context, like fact-checking, to tweets (Coleman, 2021; Conger, 2020; Feiner, 2019; Harvey, 2019; Jung et al., 2020; K. Lyons, 2021; Roth & Pickles, 2020, 2020; Twitter, 2020b, 2021a). While COVID-19–related interventions were retired after the platform's acquisition in 2022, the remaining interventions are still in effect.

In addition, some new or revised intervention strategies have been implemented, such as eliminating free API access and restricting the data usage (Saan, 2024), the "Not A Bot" policy requiring small payments for new accounts in certain regions as a deterrent to spam and bots (X, 2023), and the authenticity policy targeting coordinated inauthentic behavior that continues to be enforced (X, 2025). While the focus of our study is on Twitter (X) data, our findings and the deterrence strategies are applicable to other major platforms. For example, LinkedIn (Rockwell, 2022), Meta's platforms including Facebook, Instagram, and Threads (Facebook, n.d.; Instagram, n.d.; Meta, 2025b, 2025a), YouTube (Google, n.d.), and Bluesky (Bluesky, 2024) are also enforcing removal, reduction, and informing interventions. These examples affirm the broader relevance and adaptability of our study's framework beyond a single platform.

Over time, deterrence theory has evolved into a broader typology of deterrence strategies, each defined by the mechanisms it employs. In international relations, hard deterrence (power) refers to coercing behavior through force or penalties, while soft deterrence relies on persuasive strategies that appeal to users' values, beliefs, or understanding of consequences (Nye, 2005). Lying between these two approaches is situational deterrence (Cusson, 1993), which focuses on limiting opportunities for undesirable behavior without resorting to coercion. The removal, reduction, and informing interventions on social media can also be mapped onto these deterrence mechanisms, respectively: (1) hard deterrence, referring to coercive actions such as suspending an account, (2) situational deterrence, representing restrictive interventions such as reducing the visibility of a tweet, and (3) soft deterrence, involving persuasive actions such as labeling a tweet (Karami, 2025). Each of these interventions can produce both direct effects, such as stopping a specific action, and indirect effects, such as signaling platform expectations to a broader audience. These interventions collectively aim to reduce the spread of misinformation and enhance the quality of information exchange on the platform (Altay et al., 2023).

As shown in Table 1, studies specifically examining health misinformation mostly indicate the effectiveness of removal, reduction, or informing strategies, though some findings are contradictory.

Table 1. Studies examining the impact of the three types of interventions *(ND: Not Distinguished; NS: Not Significant; *: Significant)*

| Intervention Type | Reference | Context | User Type | Intervention Impact |
|---|---|---|---|---|
| Removal | (Allcott et al., 2019) | Diverse | ND | * |
| | (Papakyriakopoulos et al., 2020) | Health | ND | * |
| | (Jhaver et al., 2021) | Health | ND | * |
| | (Bak-Coleman et al., 2022) | Politics | ND | NS |
| | (Zhong et al., 2023) | Health | ND | NS |
| | (Broniatowski et al., 2023) | Health | ND | NS |
| | (Innes & Innes, 2023) | Health | ND | * |
| | (Horta Ribeiro, Cheng, et al., 2023) | Diverse | ND | * |
| | (McCabe et al., 2024) | Politics | Human | * |
| Reduction | (Sanderson et al., 2021) | Politics | ND | * |
| | (Vincent et al., 2022) | Diverse | ND | NS |
| | (Bak-Coleman et al., 2022) | Politics | ND | NS |
| | (Kaiser & Mayer, 2023) | Health | ND | * |
| | (Horta Ribeiro, Cheng, et al., 2023) | Diverse | ND | * |
| | (Mehta, 2023) | Politics | Human | * |
| | (Mudambi et al., 2025) | Health | Human | * |
| Informing | (Lewandowsky et al., 2012) | Diverse | ND | * |
| | (Ross et al., 2018) | Diverse | Human | NS |
| | (Kim & Dennis, 2019) | Politics | Human | * |
| | (Pennycook & Rand, 2019) | Politics | Human | * |
| | (Clayton et al., 2020) | Politics | Human | * |
| | (Papakyriakopoulos et al., 2020) | Health | ND | * |
| | (Mena, 2020) | Politics | Human | * |
| | (Porter & Wood, 2021) | Diverse | Human | * |
| | (Sanderson et al., 2021) | Politics | ND | * |
| | (Bak-Coleman et al., 2022) | Politics | ND | NS |
| | (Chipidza & Yan, 2022) | Politics | ND | NS |
| | (Bak-Coleman et al., 2022) | Politics | ND | NS |
| | (Carey et al., 2022) | Health | Human | NS |
| | (Shin et al., 2023) | Health | Human | * |
| | (Koch et al., 2023) | Climate Change | Human | * |
| | (Mehta, 2023) | Politics | Human | NS |
| | (Chuai et al., 2024) | Diverse | ND | * |
| | (Slaughter et al., 2025) | Diverse | ND | NS |
| | (Kauk et al., 2025) | Diverse | ND | NS |

For example, while Papakyriakopoulos et al. (2020) demonstrated that a content removal strategy successfully mitigated the spread of misinformation, Zhong et al. (2023) found that the same strategy was not effective. Another body of research (Table 1), which extends beyond health to domains such as politics and climate change, shows similar inconsistencies in intervention outcomes. For instance, Mena (2020) discovered that flagging political tweets could diminish the spread of false news, whereas a subsequent study by Chipidza and Yan (2022) indicated that this flagging strategy was not only ineffective but also counterproductive. These studies evaluated the efficacy of interventions (e.g., content deletion, visibility reduction, and content labeling) using survey data and real-time social media metrics. Taken together, the literature suggests that intervention effects may vary widely depending on context, user type, and platform characteristics.

Accordingly, there is a need for additional research to comprehensively assess and illuminate the effectiveness of social media platforms' measures against misinformation (Théro & Vincent, 2022). As mentioned earlier, two critical gaps exist in the current body of research. First, although both bots and humans disseminate misinformation (Cai et al., 2023; Tandoc Jr, 2019; Vosoughi et al., 2018), very few studies investigate how interventions affect these groups separately. When studies rely on aggregate misinformation measures without distinguishing between user types, the results can be misleading. If one group constitutes the majority of misinformation activity, aggregate outcomes may appear effective or ineffective simply because they reflect the dominant group's response. This undermines the goal of interventions, which is both to help humans avoid and refrain from sharing false information and to prevent bots from amplifying it. Yet current research has primarily focused on overall misinformation volume (Himelein-Wachowiak et al., 2021; Shao et al., 2017; P. Wang et al., 2018), overlooking these critical distinctions. For example, Bak-Coleman et al., (2022) simulated the spread of viral misinformation on Twitter during the 2020 U.S. election. Although the study assessed the effectiveness of interventions in reducing overall misinformation, its simplified approach did not differentiate between bots and human users, leaving unanswered questions about how each group responds to platform enforcement.

Second, the existing research mainly emphasizes the immediate (short-term) deterrence effects of interventions (Jhaver et al., 2021; Seering et al., 2017), while overlooking their sustained (long-term) impacts and how interventions shape the longer-term dynamics of misinformation on social media platforms (Bak-Coleman et al., 2022). For example, Jhaver et al. (2021a) found that the removal strategy was effective in reducing activity and toxicity levels among supporters in the short term. However, focusing solely on immediate outcomes is problematic because such evaluations may overlook the evolving nature of misinformation and whether reductions are maintained (Roozenbeek et al., 2020). Understanding the sustained effects of misinformation-combating interventions is essential for determining whether they lead to durable reductions in misinformation prevalence rather than temporary declines. This study seeks to address these two gaps by examining the long-term deterrence effect of social media interventions on bot and human accounts engaged in propagating health misinformation.

## 3- Hypothesis development

In this research, deterrence theory serves as the theoretical lens for understanding the influence of social media intervention strategies on the dissemination of misinformation.

### *3.1 Bots and deterrence interventions*

Removal, reduction, and informing interventions could exhibit a sustained deterrence effect on the number of bots spreading health misinformation via amplification (e.g., retweeting) or boundary spanning (e.g., tweeting) to not only their followers but also non-followers (Salge et al., 2022). Sustained deterrence is fostered through ongoing enforcement and the visibility of these actions, which cultivate a long-term environment of compliance and deter potential violators. Reports of social media platforms, such as Twitter biannual reports (Twitter, 2021b), indicate that it is unlikely to remove all bots. Through consistent reinforcement, the sustained deterrence effect becomes apparent as social media interventions create an enduring hostile environment for bots and their creators.

Bots often amplify their efforts by interacting with other bots (Khaund et al., 2021) and creating networks to propagate misinformation (J. Zhang et al., 2016). By targeting bots, these interventions disrupt the cycle of bot creations. Sustained deterrence can contribute to a lasting reduction in the number of bots engaged in spreading health misinformation in the following ways. The removal intervention addresses the problem at its source by eliminating bot accounts, which can dismantle coordinated efforts and reduce the overall volume of false content. The reduction intervention limits the visibility of misinformation itself, which indirectly affects bots that rely on tracking and resharing trending or popular posts. When such content is suppressed or hidden, bots have less material to amplify, decreasing their usefulness and often leading to inactivity or less interest for developers to create new bots. When their posts reach fewer people and the sources they rely on become less accessible, the bot developers may lose motivation to remain active or create new bots.

The informing intervention can help flag bot behavior patterns, enabling detection systems to identify additional bots involved in spreading misinformation. Educational efforts complement these strategies by alerting bot operators to the consequences of sharing misinformation, reinforcing the perceived risks of such actions. The heightened risk of detection and punishment likely makes bot creators less inclined to continue such activities. Together, these approaches interfere with the mechanisms bots rely on, restrict the spread of harmful content, and strengthen detection efforts over time. Against this backdrop, we expect these interventions to exert a sustained deterrence effect on the number of bots. Thus, we posit:

***H1: Social media interventions have sustained deterrence effects on the number of bots spreading health misinformation.***

> ***H1a: The removal intervention strategy has a sustained deterrence effect on the number of bots spreading health misinformation.***
> ***H1b: The reduction intervention strategy has a sustained deterrence effect on the number of bots spreading health misinformation.***
> ***H1c: The informing intervention strategy has a sustained deterrence effect on the number of bots spreading health misinformation.***

Bots, as automated systems, can adapt their activities in real time (Chu et al., 2012; Scerbo, 2007). In principle one might expect that when the number of bots decreases, the remaining bots could compensate by increasing their activity. However, such abnormally high activity levels are easily detectable and tend to trigger automated enforcement, reducing the likelihood that compensation of this kind can occur at scale (Luceri et al., 2021). Given these constraints, a reduction in the number of bots (as posited in H1) is far more likely to produce a corresponding decrease in the overall volume of misinformation rather than being offset by intensified activity among the remaining bots.

The sustained impact of removal, reduction, and informing interventions on bots is observable through two primary mechanisms. First, bots form networks by interacting with one another, creating interconnected systems that can propagate misinformation more effectively (Khaund et al., 2021; J. Zhang et al., 2016). Social media interventions disrupt these networks by decreasing the availability, visibility, influence, and reach of bots on the platform, thereby impeding the dissemination of false or misleading information across both short- and long-term periods (Czerniak et al., 2023). Second, the increased risk of punishment can deter undesirable behaviors by signaling the consequences of spreading misinformation. These punitive measures serve as clear signals to bot operators about the consequences of spreading misinformation, thus diminishing the likelihood of continued activity. Therefore, the long-lasting deterrence effects are expected to reduce the overall volume of bot-generated misinformation on the platform. Hence, we hypothesize:

***H2: Social media interventions have sustained deterrence effects on the amount of health misinformation content spread by bots.***

> ***H2a: The removal intervention strategy has a sustained deterrence effect on the level of bot activities related to health misinformation.***
> ***H2b: The reduction intervention strategy has a sustained deterrence effect on the level of bot activities related to health misinformation.***
> ***H2c: The informing intervention strategy has a sustained deterrence effect on the level of bot activities related to health misinformation.***

### *3.2 Humans and deterrence interventions*

There are human users who intentionally spread misinformation (Broniatowski et al., 2018; Katzowitz, 2018; Srikanth, 2021) and act as super-spreaders of misinformation (Himelein-Wachowiak et al., 2021). Given this, social media interventions can exert sustained effects on the number of humans sharing health misinformation. Real-time interventions of removal, reduction, and informing target users engaged in the distribution of health misinformation. These consequences can serve as a sustained deterrent, alerting users to the potential repercussions of spreading false health narratives and dissuading them from further propagation. This form of deterrence resonates with users who have previously disseminated misinformation and faced interventions, as well as those who have not, and even with potential disseminators of false content (Innes & Innes, 2023). Two key mechanisms likely drive this sustained deterrence. First, the visibility of interventions establishes a clear understanding of the risks associated with spreading misinformation. This sets a behavioral benchmark for both involved and uninvolved users and can alter user behavior by embedding a heightened awareness of the consequences of spreading false information (Jhaver et al., 2021). Second, according to the validity effect principle, misinformation, when repeatedly encountered, is more likely to be perceived as true (Boehm, 1994). Moreover, bots exhibit high activity levels in sharing misinformation (Himelein-Wachowiak et al., 2021), thereby increasing the likelihood of human exposure to and belief in false information (Ross et al., 2019). By consistently intervening with bots and humans, platforms break this cycle, reducing the frequency of exposure to false narratives and indirectly deterring human users from sharing misinformation. This not only curtails the influence of misinformation but also fosters an environment where reliable information thrives over time. In short, to reduce the number of human users who share misinformation, platforms rely on removal, reduction, and informing strategies.

Removal is the most direct method, involving the suspension or permanent banning of individuals who repeatedly violate policies, spread false information, or coordinate deceptive campaigns. This approach eliminates their presence entirely and disrupts larger networks of misinformation. Reduction also plays an important role by limiting access to and visibility of misinformation, making it harder for these users to find content to reshare. When their posts reach fewer people and the sources they rely on become less accessible, many lose interest or motivation to remain active. Informing strategies, including the use of warning labels or context indicators, can damage the credibility of individuals who frequently share misleading content. As their reputations decline and their influence weakens, some choose to leave the platform altogether. Together, these interventions not only disrupt harmful activities but also reduce the number of people engaged in spreading misinformation. Therefore, we posit:

***H3: Social media interventions have sustained deterrence effects on the number of humans spreading health misinformation.***

> ***H3a: The removal intervention strategy has a sustained deterrence effect on the number of humans spreading health misinformation.***
>
> ***H3b: The reduction intervention strategy has a sustained deterrence effect on the number of humans spreading health misinformation.***
>
> ***H3c: The informing intervention strategy has a sustained deterrence effect on the number of humans spreading health misinformation.***

Social media interventions can deliver enduring reductions in the spread of health misinformation by individuals. Their effectiveness stems from their capacity for sustained deterrence, especially in mitigating the volume of health misinformation shared by humans. Anchored in the principle of direct deterrence, these interventions directly address and curtail the amount of false content (Horta Ribeiro, Cheng, et al., 2023). When users see their content being flagged or removed, they are more likely to reconsider their information-sharing behaviour, resulting in a reduced spread of health misinformation (Aghajari et al., 2023).

The enduring impact of social media interventions, referred to as sustained deterrence, is realized in their ability to consistently reduce the quantity of misinformation shared by humans over time. This effect is deeply rooted in the concept of indirect deterrence. The visibility and consistency of the interventions serve as a constant reminder of the risks associated with sharing large volumes of unverified content (Jhaver et al., 2021). Previous research suggests that frequent exposure to false news increases the probability of individuals' acceptance of it (Del Vicario et al., 2016), engagement with misleading content (Beauvais, 2022; Bessi et al., 2014), and susceptibility to such news (Bessi et al., 2015; Del Vicario et al., 2016). Over time, these interventions not only limit the spread of individual pieces of misinformation but also reduce the overall exposure to misleading health content in the digital ecosystem. These interventions can impact not only the individuals who are actively sharing misinformation but also those who might potentially share it by reducing their exposure to such false information.

Building on this logic, each of the three intervention types contributes to sustained deterrence through distinct mechanisms. The removal intervention directly targets harmful actions by deleting misleading posts or users, comments, or shares that violate content policies, cutting off their ability to spread. Reduction interventions limit the visibility and reach of misinformation by downranking or demoting such content in feeds and search results, making it less likely to gain attention or go viral. Informing interventions, through warning labels or contextual cues, diminish the credibility and persuasiveness of misleading posts, reducing users' willingness to reshare or amplify them. These interventions discourage users from repeating the behavior when their efforts receive little engagement. The interventions not only reduce misinformation-related actions but also foster more cautious posting behavior over time. Therefore, we anticipate a sustained decline in human activities associated with misinformation at the content level. Thus, we hypothesize:

***H4: Social media interventions have sustained deterrence effects on the amount of health misinformation content spread by humans.***

> ***H4a: The removal intervention strategy has a sustained deterrence effect on the level of human activities related to misinformation.***
>
> ***H4b: The reduction intervention strategy has a sustained deterrence effect on the level of human activities related to misinformation.***
>
> ***H4c: The informing intervention strategy has a sustained deterrence effect on the level of human activities related to misinformation.***

## 4. Method

The data for this study were collected from Twitter, a platform with more than 500 million monthly active users (Shewale, 2023), generating a staggering 500 million tweets daily, where the most active country on Twitter is the U.S., with more than 80 million users (Kemp, 2023). Twitter initiated a series of interventions (Crowell, 2017), which served as a focal point for addressing our research questions. To examine the deterrence effects outlined in our hypotheses, we conducted a comparative analysis of the frequency of bots and humans, along with the corresponding content, before and after the implementation of Twitter interventions. Studying the effects of social media interventions on health issues is vital because health is a fundamental and enduring human concern, affecting all aspects of life and society. We specifically focused on *cancer misinformation*, recognizing the growing number of cancer patients and the consequential impact of misinformation in that domain[2]. In addition, malicious actors often function as networks with diverse interests, not limited to a single topic (J. Zhang et al., 2016). For instance, analysis showed that Russian trolls active during 2012–2018 engaged not only in political discussions but also extensively in health-related topics such as diet and cancer (Karami et al., 2021). In addition, misinformation regarding a specific health issue, such as COVID-19, may also be linked to misinformation about other health conditions, such as cancer (Kornides et al., 2023). This interconnection involving the network of malicious actors and their activities across various health topics suggests that our findings are not only applicable to a topic like cancer but also highly generalizable to other health contexts. A summary of the steps taken to address the research questions and test the hypotheses is presented in Table 2 and explained below.

---

[2] The National Cancer Institute (NCI) projects a surge in new annual cancer cases to 29.5 million and cancer-related deaths to 16.4 million by 2040 (NCI, 2020). Moreover, the financial burden of cancer treatment in the U.S. exceeded $21 billion in 2019 (NCI, 2020).

Table 2. A Summary of Data Collection and Analysis Procedure and Outputs

| Step | Task(s) | Output |
|---|---|---|
| **Misinformation Detection** | • Identify instances of misinformation based on fact-checking sites<br>• Remove duplicates | • 81 unique claims, i.e., instances of cancer misinformation, were identified. |
| **Social Media Data Collection** | • Develop queries to collect tweets related to the 81 false claims that potentially supported those claims. | • 268,280 tweets were collected. |
| **Machine Learning Classification** | • Develop and train a classifier based on coding 2000 tweets to identify tweets that supported cancer misinformation.<br>• Use the classifier to label the rest of the 266,280 tweets. | • 68.05% of the collected tweets were supportive of cancer misinformation. |
| **Bot Score Calculation** | • Distinguish humans and bots using bot scores generated by the tool Botometer. | • 51.5% of the accounts posting supportive tweets were humans, accounting for 40.9% of the activity, while 48.5% were bots, accounting for 59.1% of the activity. |
| **Statistical Analysis (Hypothesis Testing)** | • Perform ARIMA models to test the hypotheses | • Sustained effects of interventions on bots and humans were revealed (*explained in the results section*) |

Our results are complemented by post hoc analyses based on analyzing the number of all users and their activities without distinguishing between bots and humans, helping to illustrate how the absence of this distinction could influence the overall findings.

### *4.1 Misinformation detection*

To identify cancer-related misinformation, we used the Google Fact Check API (Application Programming Interfaces), accessible at https://toolbox.google.com/factcheck/apis. This tool verifies information through independent fact-checking organizations. We narrowed our query to English-language information containing the keyword “cancer” and published between 2011 and 2021. Various features were automatically extracted, with a particular emphasis on text, date, and a rating assessment of the truthfulness of information. Our study focused on untrue claims. We removed duplicate entries. From the Google Fact Check database, we collected a total of 111 claims, of which 84 claims represented unique content. We excluded true claims, including those for which the Google Fact Check did not confirm falsehood. This results in 81 instances of cancer misinformation used as queries for the social media data collection.

### *4.2 Social media data collection*

Twitter data has been used to study a wide range of phenomena, such as analyzing data related to disasters (Son et al., 2019), health (Karami et al., 2025), and politics (Georgiadou et al., 2020). To obtain Twitter data, we developed a query for each of the misinformation claims detected by Google Fact Check. For example, the query *“(marijuana OR cannabis OR weed OR pot) AND (cure OR kill OR treat) AND (cancer)*” was used for the following misinformation claim: “*Marijuana is non-addictive, cures cancer, and grows brain cells.*” Using these queries, we collected English tweets through the Brandwatch service, which is an official Twitter partner (now X)[3]. In total, we collected 268,280 posts, including tweets, retweets, replies, and threads posted by 150,171 unique users in the U.S. between 2011 and 2021.

### *4.3 Machine learning classification*

We developed a classifier employing state-of-the-art machine learning methods to identify tweets supporting cancer misinformation from the 268,280 potentially relevant tweets. An initial screening of a random sample of tweets resulted in categorizing them into three groups: *supportive*, *irrelevant*, and *other*. The *supportive* tweets were those that endorsed cancer misinformation. The *irrelevant* group comprised tweets containing query-related keywords but not actually related to cancer misinformation. The group labeled as *other* included tweets that were relevant but did not endorse cancer misinformation, such as humorous comments.

Out of the 268,280 tweets, one coder annotated the 2000 tweets for the training dataset, ensuring a minimum of 5 tweets per claim. Among these, 1249 (62.45%) were labeled as supportive, 452 (22.60%) as irrelevant, and 299 (14.95%) as other. Maintaining class proportions, we randomly selected 63 supportive tweets, 22 irrelevant tweets, and 15 other tweets (100 in total) for coding by a second coder. The agreement between the two coders was substantial, as indicated by Cohen’s $\kappa = 0.73$, indicating substantial agreement between the coders (McHugh, 2012).

Our classifier employed unigrams, bigrams, and trigrams to predict whether a tweet was supportive, irrelevant, or other. This involved converting tweets into sequences of one, two, and three terms. We applied five high-performance classification algorithms, including Bayesian Network (BayesNet), Support Vector Machines (SVM), Linear Logistic Regression, Random Forest, and Convolutional Neural Network (CNN), and compared their performance to identify the best-performing algorithm for this study (Fernández-Delgado et al., 2014; Pham et al., 2017; X. Wu et al., 2008; C. Zhang et al., 2017). To enhance the efficiency and effectiveness of the algorithms, we used dimensionality reduction to simplify the data without losing critical information. Evaluation of the algorithms involved a five-fold cross-validation approach, wherein tweets were divided into five subsets, with one set aside for testing and the remaining four used for training. Metrics such as accuracy, F-measure, precision, and recall were employed to compare the algorithms and identify the most effective features. As a result, each tweet was labeled as supportive, irrelevant, or other. The performance of the five classification algorithms was then compared. Through Chi-Squared ($\chi^2$) test experiments assessing the effects of varying numbers of features on classifier accuracy, the Linear Logistic Regression model demonstrated superior performance across different feature sets. Our classifier achieved an accuracy of 79.75%, F-measure of 0.788, recall of 0.798, and precision of 0.791. We also performed a t-test to compare the performance of the Linear Logistic Regression model to the ZeroR baseline classifier, predicting the most frequently occurring category or class in the dataset without considering any predictors (Lane et al., 2012). The results indicated that the improvement of the Linear Logistic Regression model over the baseline was statistically significant ($p$-value < 0.05). The classifier was then applied to the remaining 266,280 unlabeled tweets, yielding 181,323 (68.05%) supportive, 58,324 (21.91%) irrelevant, and 26,633 (10.04%) other tweets (Table 3). Figure 1 shows the machine learning classification process from feature extraction to labeling unlabeled tweets. Furthermore, out of the 150,171 unique users, 96,836 (64.5%) were involved in posting tweets belonging to the supportive class, which is the focus of the remainder of this study.

---

[3] https://www.brandwatch.com/partnerships/x/

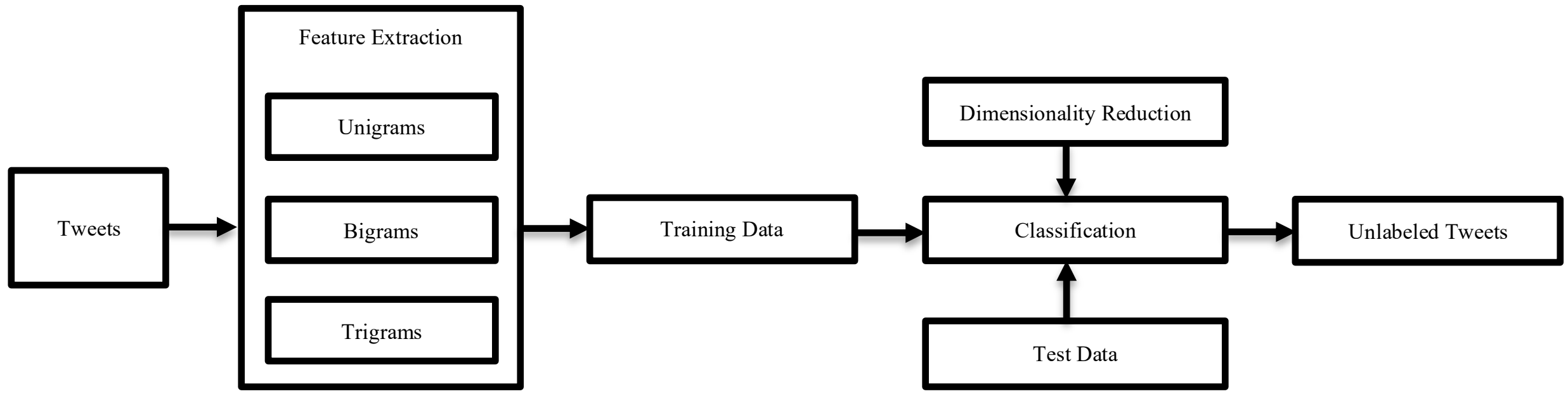

Figure 1: Machine learning classification process.

Table 3. Basic Statistics of Categorizers.

| Data | Total | Relevant | | Irrelevant |
|---|---|---|---|---|
| | | Supportive | Other | |
| **Training Data** | 2000 | 1249 | 299 | 452 |
| **Unlabeled Data** | 264,280 | 180,074 | 26,334 | 57,872 |
| **Total** | 266,280 | 181,323 | 26,633 | 58,324 |

#### *4.4 Bot score measurement*

To categorize users as humans or bots, we used Botometer, a tool that employs a supervised machine learning model trained on a large dataset with over 1000 features (Yang et al., 2019) to assess the likelihood of an account being a bot. The features used by Botometer are from six groups: user metadata (e.g., number of followers), friends (e.g., distribution of friends), content (e.g., number of words in a tweet), sentiment (e.g., number of negative emotions in tweets), network (e.g., network density), and timing (e.g., time interval between consecutive tweets) (Varol et al., 2017). Using these features, Botometer assigned each account supporting cancer misinformation a score from 1 (likely human) to 5 (likely bot). Among the 96,836 accounts endorsing cancer misinformation, 66.9% received scores between 1 and 5, while the rest were suspended (14.2%), non-existent (13.3%), protected (4.5%), or had their tweets removed (1.1%).

Thresholds for bot scores vary in previous studies, ranging from 2.5 (Bessi & Ferrara, 2016) to 4 (Broniatowski et al., 2018). A threshold of 2.5 is common (Bessi & Ferrara, 2016) but increases the risk of false positives (Gallwitz & Kreil, 2022), while higher thresholds may increase the risk of false negatives. To address this, we computed means for bots and humans, including their tweets and retweets, across thresholds from 2.5 to 4 in increments of 0.1. On average, humans constituted 51.5% of users and contributed 40.9% of activities, while bots represented 48.5% of users but were involved in 59.1% of activities.

#### *4.5 Statistical analysis*

We employed interrupted time-series analysis (ITSA) (Nagin, 1998) to test our hypotheses. ITSA allows for multiple measurements before and after an intervention, enabling the exploration of longitudinal trends in intervention effects. We used the autoregressive integrated moving average (ARIMA) model (Box & Jenkins, 1976), which effectively isolates the impact of the intervention from overall trends, long-term patterns, outliers, seasonal variations, and sequential dependencies within the data (McCleary et al., 1980). Seasonal patterns in social media discussions may be influenced by various factors, such as awareness campaigns dedicated to different issues (AACR, 2023; HHS, 2023) and celebrity endorsements (e.g., skin cancer (Vasconcelos Silva et al., 2020)), which can also contribute to fluctuations in social media conversations about cancer. To account for these seasonal effects, we applied a seasonal ARIMA model.

This study focused on the sustained effect (step-change or level-shift), which captures the lasting change in the outcome variable (e.g., number of bots) that occurs once an intervention is implemented and persists for as long as the intervention remains in place. In this specification, the indicator variable ($S_t$) takes the value 0 when the intervention is absent (before $T_0$) and switches to 1 when the intervention is present (after $T_0$) (Schaffer et al., 2021):

$$S_t = \begin{cases} 0, if\ t < T_0 \\ 1, if\ t \geq T_0 \end{cases}$$

To determine the initiation date of each intervention, we reviewed Twitter's official blog announcements and identified the initial posts introducing the removal, reduction, and informing interventions, published in February 2017 (Ho, 2017), June 2018 (Roth & Harvey, 2018), and September 2020 (Twitter, 2020a), respectively. The months following each announcement were treated as the post-intervention period.

To evaluate the intervention's effect, we examined the coefficients associated with the intervention parameters and assessed their statistical significance. In this study, these coefficients represent the change in the average monthly volume of misinformation posted on social media following the start point of each intervention. Additional methodological details are provided in Appendix A.

While seasonal ARIMA can manage seasonal variations in social media discussions, we further evaluated the robustness of our findings by analyzing Google search trends for the term "Cancer" to rule out the possibility that external events, either on Twitter or in the broader information environment, were driving shifts in cancer-related (mis)information. We examined search interest for "Cancer" before and after the initial introduction of Twitter's intervention measures (Figure 2) and found no significant difference between the two periods ($p > 0.05$). In addition, trend analysis (Hamilton, 2020) showed no significant upward or downward trend from July 2011 through December 2021, indicating stable search behavior for "Cancer" over a decade.

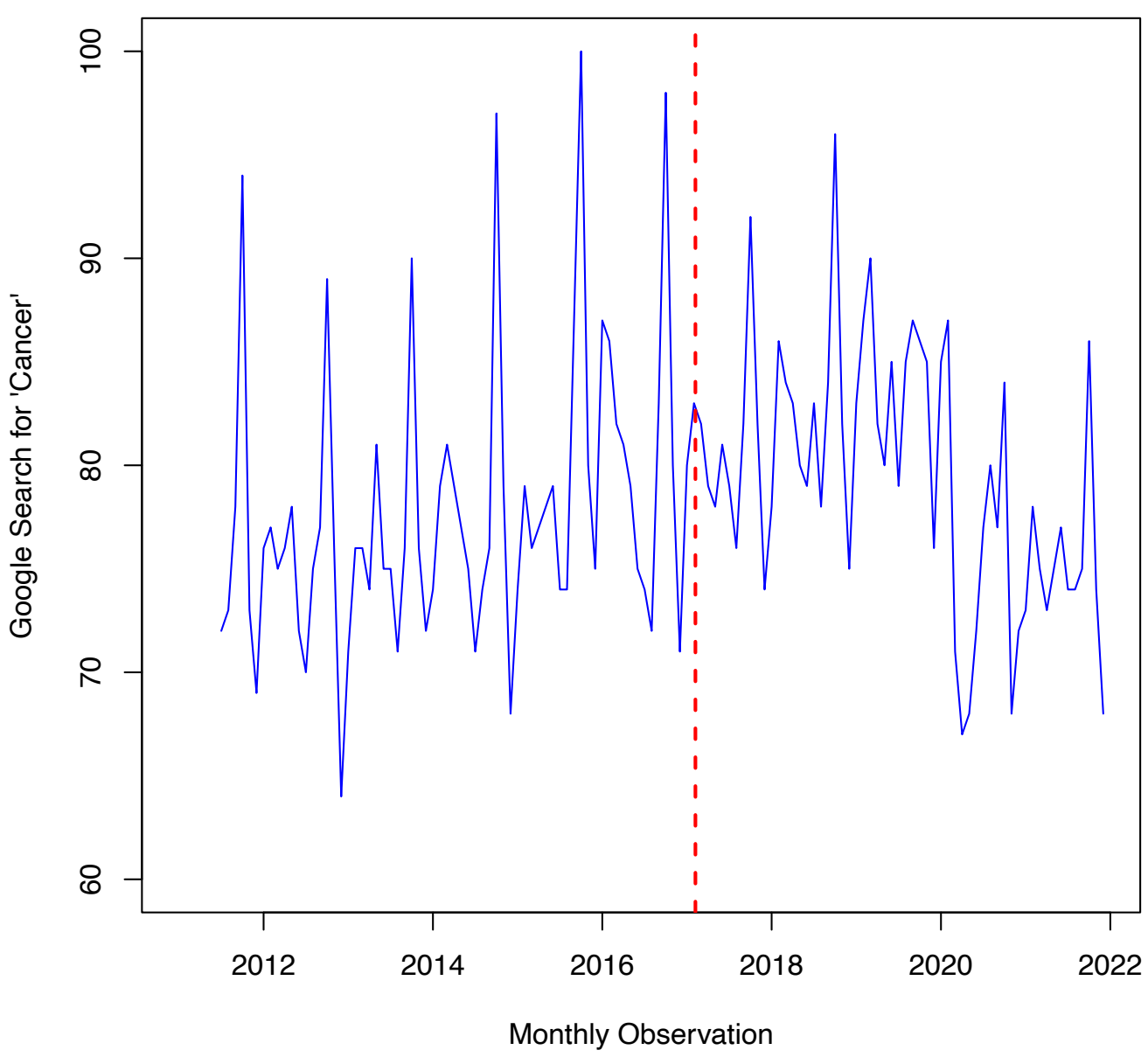

Figure 2. Monthly Google search interest for "Cancer" with a vertical line marking the introduction of Twitter's intervention measures.

## 5- Results

Table 4 presents the sustained effects of the interventions on bot presence. The deterrence parameters for the removal, reduction, and informing strategies were -0.804, -0.892, and -0.554, respectively, all statistically significant (p-value < 0.05). The negative coefficients indicate that each intervention was effective in reducing the presence of bots on Twitter. Hence, H1a, H1b, and H1c were supported. The antilog of the removal deterrence coefficient ($10^{-0.708}$) corresponds to an 84% decrease in the average presence of bots following implementation. Similarly, the reduction and informing interventions corresponded to decreases of 87% and 72%, respectively. Ljung-Box tests, used for assessing autocorrelation in the residuals, yielded p-values well above the 0.05 threshold across various degrees of freedom (ranging from 3 to 20), confirming that the model residuals did not show significant autocorrelation and the model was adequate.

Table 4. Summary of ARIMA intervention model estimation for bot users across the three intervention strategies (H1).

| Model Estimation: Bot Users | | | | |
|---|---|---|---|---|
| Variable | Estimate | SE | T | *p-value* |
| Removal | -0.804 | 0.250 | -3.216 | 0.001* |
| Reduction | -0.892 | 0.256 | -3.483 | 0.000* |
| Informing | -0.554 | 0.274 | -2.024 | 0.043* |
| **Model Assessment** | | | | |
| Stationary $R^2$ | 0.187 | | | |
| RMSE | 0.410 | | | |
| MAE | 0.299 | | | |
| Ljung-Box statistics | X-squared = 1.456, df = 3, p-value = 0.692 | | | |
| | X-squared = 1.734, df = 4, p-value = 0.784 | | | |
| | X-squared = 2.652, df = 5, p-value = 0.753 | | | |
| | X-squared = 5.828, df = 10, p-value = 0.829 | | | |
| | X-squared = 6.688, df = 12, p-value = 0.877 | | | |
| | X-squared = 15.593, df = 20, p-value = 0.741 | | | |

Table 5 summarizes the sustained effects of the interventions on bot activities. While the informing intervention was not significant, the removal and reduction strategies showed significant deterrence parameters of -1.762 and -0.859, respectively (p-value < 0.05), supporting H2a and H2b, but not H3c. The antilog of the removal coefficient ($10^{-1.762}$) corresponded to a 98% reduction in bot activities, whereas the reduction strategy yielded an 86% decrease. The Ljung-Box statistic again showed p-values significantly above 0.05, indicating no residual autocorrelation and confirming that the model was adequate. Overall, these results suggest that the removal and reduction interventions were highly effective in the reduction of bot activities.

Table 5. Summary of ARIMA intervention model estimation for bot activities across the three intervention strategies (H2).

| Model Estimation: Bot Activities | | | | |
|---|---|---|---|---|
| Variable | Estimate | SE | t | *p-value* |
| Removal | -1.762 | 0.349 | -5.051 | 0.000* |
| Reduction | -0.859 | 0.346 | -2.481 | 0.013* |
| Informing | -0.176 | 0.356 | -0.496 | 0.620 |
| **Model Assessment** | | | | |
| Stationary $R^2$ | 0.248 | | | |
| RMSE | 0.439 | | | |
| MAE | 0.315 | | | |
| Ljung-Box statistics | X-squared = 0.505, df = 3, p-value = 0.918 | | | |
| | X-squared = 0.548, df = 4, p-value = 0.969 | | | |
| | X-squared = 1.175, df = 5, p-value = 0.947 | | | |
| | X-squared = 6.420, df = 10, p-value = 0.779 | | | |
| | X-squared = 6.575, df = 12, p-value = 0.884 | | | |
| | X-squared = 16.262, df = 20, p-value = 0.700 | | | |

Table 6 reports the ARIMA results estimating the sustained effects of the interventions on the number of human users sharing misinformation. None of the interventions produced statistically significant changes in the number of humans sharing misinformation; therefore, H3a, H3b, and H3c were not supported. Model diagnostics using the Ljung-Box statistic were satisfactory, showing no significant autocorrelation at the 0.05 level across different degrees of freedom.

Table 6. Summary of ARIMA intervention model estimation for human users across the three intervention strategies (H3).

| **Model Estimation: Human Users** | | | | |
|---|---|---|---|---|
| Variable | Estimate | SE | t | *p-value* |
| Removal | 0.233 | 0.275 | 0.845 | 0.398 |
| Reduction | 0.042 | 0.263 | 0.160 | 0.873 |
| Informing | -0.280 | 0.280 | -1.000 | 0.317 |
| **Model Assessment** | | | | |
| Stationary $R^2$ | 0.366 | | | |
| RMSE | 0.392 | | | |
| MAE | 0.296 | | | |
| Ljung-Box statistics | X-squared = 0.518, df = 3, p-value = 0.915 | | | |
| | X-squared = 0.566, df = 4, p-value = 0.968 | | | |
| | X-squared = 1.135, df = 5, p-value = 0.951 | | | |
| | X-squared = 1.198, df = 10, p-value = 0.999 | | | |
| | X-squared = 1.363, df = 12, p-value = 0.999 | | | |
| | X-squared = 8.220, df = 20, p-value = 0.990 | | | |

Table 7 summarizes the sustained effects of the interventions on human activities related to misinformation. The estimated coefficients for the removal intervention, reduction, and informing interventions were not significant. The Ljung-Box statistic showed no autocorrelation in the residuals, supporting model adequacy. Overall, the interventions did not significantly alter human behavior related to misinformation posting over time, and thus H4a, H4b, and H4c were not supported.

Table 7. Summary of ARIMA intervention model estimation for human activities across the three intervention strategies (H4).

| **Model Estimation: Human Activities** | | | | |
|---|---|---|---|---|
| Variable | Estimate | SE | T | *p-value* |
| Removal | 0.248 | 0.269 | 0.921 | 0.357 |
| Reduction | -0.003 | 0.261 | -0.011 | 0.991 |
| Informing | -0.304 | 0.279 | -1.092 | 0.275 |
| **Model Assessment** | | | | |
| Stationary $R^2$ | 0.373 | | | |
| RMSE | 0.379 | | | |
| MAE | 0.290 | | | |
| Ljung-Box statistics | X-squared = 0.758, df = 3, p-value = 0.860 | | | |
| | X-squared = 0.916, df = 4, p-value = 0.922 | | | |
| | X-squared = 1.399, df = 5, p-value = 0.924 | | | |
| | X-squared = 2.513, df = 10, p-value = 0.991 | | | |

| | |
|---|---|
| | X-squared = 2.682, df = 12, p-value = 0.997<br>X-squared = 9.084, df = 20, p-value = 0.982 |

These statistical findings align with the visual patterns shown in Figure 3. Figures 3a and 3b show clear declines in bot presence and bot activities following the implementation of removal, reduction, and informing strategies, and these declines persist over time, consistent with support for H1a–H1c and H2a–H2b. In contrast, Figures 3c and 3d reveal no consistently declining pattern for human users.

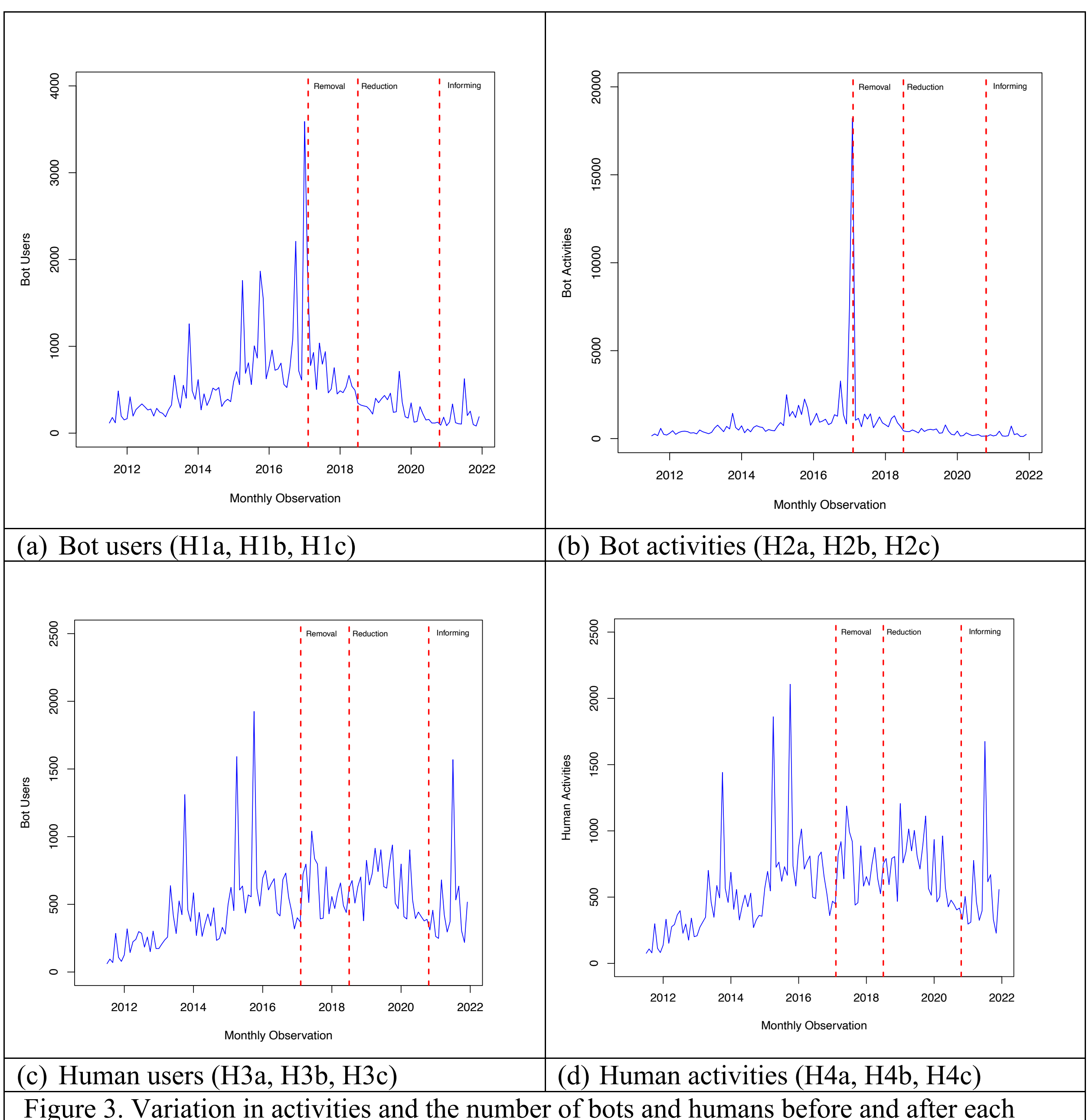

(a) Bot users (H1a, H1b, H1c)

(b) Bot activities (H2a, H2b, H2c)

(c) Human users (H3a, H3b, H3c)

(d) Human activities (H4a, H4b, H4c)

Figure 3. Variation in activities and the number of bots and humans before and after each intervention's start point marked with vertical line.

Both the number of humans sharing misinformation (H3a–H3c) and the volume of their misinformation activities (H4a–H4c) remained relatively stable throughout the observation period, reinforcing the finding that the interventions were far less effective in producing sustained changes in either the number of humans sharing misinformation or the volume of their misinformation activities.

## 6- Discussion

This longitudinal research examined the sustained deterrence effects of social media interventions on bot and human accounts engaged in disseminating health misinformation. To do so, this study leveraged a combination of computational techniques, including machine learning algorithms and statistical models, to rigorously test the hypotheses outlined in Table 8.

Table 8. Summary of Hypothesis Testing Results

| Hypothesis | Hypothesized Effect | Result |
|---|---|---|
| H1a | The removal intervention strategy has a sustained deterrence effect on the number of bot users spreading health misinformation. | Supported |
| H1b | The reduction intervention strategy has a sustained deterrence effect on the number of bot users spreading health misinformation. | Supported |
| H1c | The informing intervention strategy has a sustained deterrence effect on the number of bot users spreading health misinformation. | Supported |
| H2a | The removal intervention strategy has a sustained deterrence effect on the level of bot activities related to health misinformation. | Supported |
| H2b | The reduction intervention strategy has a sustained deterrence effect on the level of bot activities related to misinformation. | Supported |
| H2c | The informing intervention strategy has a sustained deterrence effect on the level of bot activities related to health misinformation. | Not supported |
| H3a | The removal intervention strategy has a sustained deterrence effect on the number of human users spreading health misinformation. | Not supported |
| H3b | The reduction intervention strategy has a sustained deterrence effect on the number of human users spreading health misinformation. | Not supported |
| H3c | The informing intervention strategy has a sustained deterrence effect on the number of human users spreading health misinformation. | Not supported |
| H4a | The removal intervention strategy has a sustained deterrence effect on the level of human activities related to misinformation. | Not supported |
| H4b | The reduction intervention strategy has a sustained deterrence effect on the level of human activities related to misinformation. | Not supported |
| H4c | The informing intervention strategy has a sustained deterrence effect on the level of human activities related to misinformation. | Not supported |

While the results mostly confirmed the effectiveness of interventions on bots, they did not demonstrate comparable deterrence effects for human users, regardless of the intervention strategy. Several factors help explain why interventions struggle to deter humans over time. First, prolonged exposure to health misinformation before interventions are implemented can entrench false beliefs, reducing the effectiveness of subsequent corrective actions (Moravec et al., 2019); this challenge is amplified when interventions occur with delays rather than immediately, as timely action is critical for reducing misinformation at scale (Truong et al., 2025). Second, even after false information is corrected, the belief echoes effect (Thorson, 2016) allows misinformation to continue shaping attitudes, which limits the long-term impact of platform responses. Third, cognitive tendencies such as the truth-default theory (i.e., believing that others tell the truth more often) (Levine, 2014), naïve realism (i.e., believing that one's current views are the only correct views) (Shu et al., 2020), and confirmation bias (i.e., processing information that aligns with one's current beliefs) (Schuetz et al., 2021; Shu et al., 2020) further undermine individuals' ability to critically evaluate and resist misinformation. Fourth, interventions (e.g., suspending malicious actors) may redirect rather than eliminate human participation, as suspended or blocked users often migrate and repost content on other platforms (Horta Ribeiro, Hosseinmardi, et al., 2023), as seen when blocked comments on Twitter reappeared on Facebook, Instagram, and Reddit (Sanderson et al., 2021). Collectively, these factors suggest that even when interventions are implemented, lingering false beliefs, cognitive biases, and cross-platform migration reduce the likelihood that humans will decrease their misinformation-sharing behavior.

### *6.1 Theoretical contributions*

This research makes several significant contributions to the literature on misinformation and social media interventions. First, building on the approach of explaining phenomena through theory, this study extends deterrence theory (Stafford & Warr, 1993) by contextualizing its three common forms, namely hard deterrence (coercive), situational deterrence (restrictive), and soft deterrence (persuasive), within the domain of social media misinformation and interventions. Although these forms have traditionally been applied in areas such as criminal justice (Tomlinson, 2016), cybersecurity (D'arcy & Herath, 2011), and organizational misconduct (Hensel & Kacprzak, 2021), they have not been systematically linked to platform-level intervention strategies. By mapping them onto the three major social media interventions, removal (hard), reduction (situational), and informing (soft), this study offers a clear conceptual foundation for understanding how interventions operate as distinct deterrence mechanisms and for evaluating their sustained effects in digital environments. Importantly, our empirical findings demonstrate that the three forms of deterrence differ not only conceptually but also in their practical effectiveness when operationalized as platform interventions. By adopting deterrence theory in empirical observations from a real-world misinformation context, this study advances a more nuanced understanding of how different intervention strategies function as distinct deterrence mechanisms on social media platforms.

Second, this study introduces a user-type specific perspective by analytically separating bots from human users, an approach that remains largely underdeveloped in the misinformation literature. Prior studies often evaluate intervention effects on aggregated user activity (Table 1), which obscures the fundamentally different roles, motivations, and behavioral dynamics of automated and human actors. By distinguishing between these two groups, this study demonstrates that deterrence mechanisms do not operate uniformly across users. Bots show clear, substantial, and sustained reductions in both presence and activity following removal and reduction interventions, whereas humans exhibit no statistically significant decline in either participation or misinformation-sharing behavior over time. This is an important finding, as prior studies have noted that interventions were generally effective in reducing all types of users and their misinformation sharing activities on social media platforms (Bak-Coleman et al., 2022). Moreover, while prior studies suggest a positive association between bot activities and human activities (Shao et al., 2018), our results indicate that the interventions primarily succeed in reducing the number of bots and their activities, rather than humans. This result stands partially in contrast to previous research suggesting an overall positive impact of social media interventions on curbing misinformation spread (Horta Ribeiro, Cheng, et al., 2023; Jhaver et al., 2021; Papakyriakopoulos et al., 2020). To further verify this distinction, we conducted a post hoc analysis mirroring prior aggregated approaches (Appendix B). The analysis confirms the positive effect of removal interventions on user presence when bots and humans are not distinguished. However, it highlights that these results could be misleading if the goal is to specifically understand the impact on human behavior.

Third, this study advances the literature by shifting attention from immediate intervention effects, which dominate prior research, to sustained deterrence effects observed over a decade of platform activity. Most existing studies evaluate interventions shortly after implementation (Table 1), which limits understanding of whether platform actions remain effective as misinformation ecosystems evolve. By using longitudinal data and interrupted time-series analyses, this study demonstrates how the impact of removal, reduction, and informing strategies unfolds over long periods and how their effectiveness changes as actors, particularly human actors, adapt to platform policies. This long-term perspective clarifies that some interventions retain their deterrent influence, while others diminish in effectiveness, offering an essential corrective to conclusions drawn from short-term evaluations (Table 1). The findings thus contribute a temporally grounded understanding of intervention efficacy that has been largely absent from prior misinformation research.

Collectively, these contributions shift the theoretical understanding of misinformation interventions from a static, user-agnostic view to a dynamic, actor-specific, and temporally grounded framework.

### *6.2 Implications for Practice*

The findings of this study offer several important practical implications for social media platforms seeking to curb the spread of health misinformation. The results suggest that social media interventions can have sustained deterrence effects on bots, effectively reducing their numbers and curbing the rapid spread of false information, which tends to propagate faster than accurate information (Vosoughi et al., 2018). This indicates that social media platforms have become increasingly proficient in utilizing algorithms to detect and remove bots. This process typically involves machine learning models that analyze user behavior, searching for signs that an account may be automated. For instance, a bot may post more frequently than a human reasonably could, or its posts might follow specific patterns. When these algorithms identify a potential bot, the platform can take immediate action, such as suspending the account (removal), limiting its visibility (reduction), or flagging its content as potentially misleading (informing). Consistent enforcement, clearer communication about enforcement practices, and transparency regarding penalties can further reinforce deterrence among automated actors. The reduction in bot activities suggests that efforts to combat the spread of health misinformation by targeting bot accounts can be effective. This not only helps prevent automated manipulation and misinformation-induced panic but also safeguards public health, promotes access to reliable and accurate information, and enhances trust and credibility in social media platforms.

However, while the interventions had discernible deterrence effects on bots, these effects were not observed in the case of human users. This highlights the need for distinct intervention strategies targeting bots and human users to combat health misinformation effectively. For bots, penalties such as suspension may be appropriate, while human users could benefit from access to reliable sources of information. Research has demonstrated that merely knowing that something is false does not necessarily prevent people from believing or sharing it, and false information can persistently influence individuals' beliefs (Bai, 2021). Therefore, a more proactive approach is needed to reduce the spread and exposure to misinformation from the outset by identifying and labeling misinformation (Varzgani et al., 2023). If people do not remember the false tag attached to misinformation or the tag was not attached at all, their minds often default to assuming accuracy (Bai, 2021). According to false tagging theory, the mental processes involved in skepticism and critical thinking are located in a different part of the brain than the processes responsible for perceiving and remembering information (Asp & Tranel, 2013). This suggests that when people encounter new information, they tend to initially treat it as true, even if only temporarily, and later assess its accuracy (Gilbert et al., 1993). Given these cognitive tendencies, combining intervention strategies (Bak-Coleman et al., 2022) with human-machine interaction and collaboration tools (Tran et al., 2023) may offer more effective pathways for curbing health misinformation among human users.

Our findings underscore the importance of distinguishing between bots and humans. Bots and humans behave differently, and without separating these groups in analysis, the results can be misleading. A reduction in overall misinformation may falsely suggest success across all users, when in reality, the decrease could be limited to bots while misinformation spread by humans continues or worsens. Differentiating user types also supports more efficient resource allocation, as platforms might otherwise over-prioritize bot-oriented interventions while neglecting strategies better suited to addressing human behavior. Targeting each group with tailored approaches ensures that platforms respond more effectively and allocate operational resources more strategically.

***6.3 Limitations and Future Work***

One study does not come without limitations, which presents opportunities for future research. First, our study primarily focused on examining the deterrence effect of Twitter interventions on both bots and humans. It is crucial to note that while common features exist among various social media platforms, such as (re)sharing and engagement mechanisms, our findings may not necessarily be universally applicable to other social media platforms. Each platform possesses unique features and moderation strategies, which can significantly influence the efficacy of interventions and their deterrent effects. Thus, future research should explore the effectiveness of interventions and deterrence strategies across different social media platforms, taking into account their specific characteristics and mechanisms.

Second, this study lacks access to sociodemographic information, such as age, gender, ethnicity, education level, and political orientation, for human users included in the dataset. As a result, we are unable to analyze how different demographic groups may respond differently to misinformation or to platform interventions. To enhance and update our understanding of this phenomenon, future research can investigate the impact of interventions on users based on sociodemographic factors and explore how bot strategies changed before and after implementing the interventions.

Third, data collection for this study was limited to misinformation claims and tweets written in English and posted in the U.S. This limitation to some extent restricts our understanding of the global landscape of misinformation, as different languages and countries have unique cultural contexts and dynamics. Future research may explore the effects of interventions on content in languages other than English that originates from foreign locations.

Fourth, large language models (LLMs), such as ChatGPT, have powerful capabilities to generate realistic text on diverse topics, such as health (Karami et al., 2024), and can be used to create misinformation with deceptive intentions (Radivojevic et al., 2024). The overarching field of machine behavior emphasizes that understanding the behavior of artificial intelligence systems is essential to control their actions, reap their benefits, and minimize their harms**.** This is particularly critical given the unprecedented prevalence of diverse algorithms, including bot algorithms on social media, which significantly influence the information citizens encounter (Rahwan et al., 2019)**.** Although LLM-powered bots are still rare on Twitter, they can sometimes be identified by self-revealing phrases like "*as an AI language model*" and "*violates OpenAI's content policy*" (Dekens, 2023; Yang & Menczer, 2024). There is a debate on how much current bot detection tools can identify bots that use LLMs for content creation (Li et al., 2023; Yang & Menczer, 2024). Humans often engage with, amplify, or imitate bot-generated content, regardless of whether that content originates from rule-based bots or LLMs systems. The emerging interdisciplinary field of machine behavior specifically frames and surveys the scientific study of intelligent machines, treating social bots and humans empirically. This field particularly emphasizes the study of human and bot behavior, which examines bots and humans in complex social environments (Rahwan et al., 2019)**.** Understanding this human-bot dynamic is crucial for designing effective public health communication strategies and misinformation interventions. This study is based on social media data collected before ChatGPT's release in November 2022 (Hu, 2023). Therefore, it is crucial to develop new detection methods for LLM-powered bots and replicate this study to assess the effectiveness of social media intervention strategies, which may face challenges in addressing these advanced bots. As such, results of those studies can be compared against our findings to understand how LLM-powered bots have changed the effectiveness of intervention mechanisms.

## 7. Conclusion

This study represents the initial endeavor to examine the immediate and sustained deterrence effects of social media interventions on both bots and humans engaged in the dissemination of health misinformation. The results revealed that social media interventions had significant and lasting deterrence effects on bots, while similar effects were not observed in the case of humans. This research carries significant theoretical and practical implications and extends the deterrence theory by offering fresh insights through the analysis of social media data. The findings can inform the development and assessment of tailored social media interventions targeting both bots and humans. By maintaining vigilance and applying moderation measures on social media platforms to combat false health information, we can strive toward a healthier and more informed society. This study stands to benefit researchers, healthcare professionals, and policymakers in their efforts to detect and analyze bot and human behaviors and investigate the impact of social media interventions on these activities, particularly concerning health-related and non-health-related domains.

**Funding**

No funding was received.

**Disclosure of Interest**

No potential conflict of interest was reported by the authors.

## References

AACR. (2023). Cancer Awareness Months. *American Association for Cancer Research (AACR)*. https://www.aacr.org/patients-caregivers/awareness-months/

Abramovaite, J., Bandyopadhyay, S., Bhattacharya, S., & Cowen, N. (2023). Classical deterrence theory revisited: An empirical analysis of Police Force Areas in England and Wales. *European Journal of Criminology*, *20*(5), 1663–1680. https://doi.org/10.1177/14773708211072415

Aghajari, Z., Baumer, E. P. S., & DiFranzo, D. (2023). Reviewing Interventions to Address Misinformation: The Need to Expand Our Vision Beyond an Individualistic Focus. *Proceedings of the ACM on Human-Computer Interaction*, *7*(CSCW1), 1–34. https://doi.org/10.1145/3579520

Allcott, H., Gentzkow, M., & Yu, C. (2019). Trends in the diffusion of misinformation on social media. *Research & Politics*, *6*(2), 205316801984855. https://doi.org/10.1177/2053168019848554

Altay, S., Berriche, M., Heuer, H., Farkas, J., & Rathje, S. (2023). A survey of expert views on misinformation: Definitions, determinants, solutions, and future of the field. *Harvard Kennedy School Misinformation Review*, *4*(4), 1–34.

Angelucci, C., Gutmann, M., & Prat, A. (2024). *Beliefs About Political News in the Run-up to an Election*. National Bureau of Economic Research. https://www.nber.org/papers/w32802

Asp, E., & Tranel, D. (2013). False tagging theory. *Principles of Frontal Lobe Function*, 383–416.

Bai, H. (2021). *Fake News Known as Fake Still Changes Beliefs and Leads to Partisan Polarization*. https://doi.org/10.31234/osf.io/v9gax

Bak-Coleman, J. B., Kennedy, I., Wack, M., Beers, A., Schafer, J. S., Spiro, E. S., Starbird, K., & West, J. D. (2022). Combining interventions to reduce the spread of viral misinformation. *Nature Human Behaviour*, *6*(10), 1372–1380.

Bauchner, H. (2019). Trust in health care. *Jama*, *321*(6), 547–547. https://doi.org/10.1001/jama.2018.20795

Beauvais, C. (2022). Fake news: Why do we believe it? *Joint Bone Spine*, *89*(4), 105371.

Beccaria, C. (2016). *On crimes and punishments*. Transaction Publishers.

Bessi, A., Coletto, M., Davidescu, G. A., Scala, A., Caldarelli, G., & Quattrociocchi, W. (2015). Science vs conspiracy: Collective narratives in the age of misinformation. *PloS One*, *10*(2), e0118093.

Bessi, A., & Ferrara, E. (2016). Social bots distort the 2016 US Presidential election online discussion. *First Monday*.

Bessi, A., Scala, A., Rossi, L., Zhang, Q., & Quattrociocchi, W. (2014). The economy of attention in the age of (mis)information. *Journal of Trust Management*, *1*(1), 12. https://doi.org/10.1186/s40493-014-0012-y

Bluesky. (2024). *Community Guidelines*. Bluesky. https://bsky.social/about/support/community-guidelines

Boehm, L. E. (1994). The validity effect: A search for mediating variables. *Personality and Social Psychology Bulletin*, *20*(3), 285–293.

Boshmaf, Y., Muslukhov, I., Beznosov, K., & Ripeanu, M. (2013). Design and analysis of a social botnet. *Computer Networks*, *57*(2), 556–578.
Box, G. E., & Jenkins, G. M. (1976). Time series analysis: Forecasting and control San Francisco. *Calif: Holden-Day*.
Brockwell, P. J., & Davis, R. A. (Eds.). (2002). State-Space Models. In *Introduction to Time Series and Forecasting* (pp. 259–316). Springer New York. https://doi.org/10.1007/0-387-21657-X_8
Broniatowski, D. A., Jamison, A. M., Qi, S., AlKulaib, L., Chen, T., Benton, A., Quinn, S. C., & Dredze, M. (2018). Weaponized health communication: Twitter bots and Russian trolls amplify the vaccine debate. *American Journal of Public Health*, *108*(10), 1378–1384.
Broniatowski, D. A., Kerchner, D., Farooq, F., Huang, X., Jamison, A. M., Dredze, M., Quinn, S. C., & Ayers, J. W. (2022). Twitter and Facebook posts about COVID-19 are less likely to spread misinformation compared to other health topics. *PloS One*, *17*(1), e0261768. https://doi.org/10.1371/journal.pone.0261768
Broniatowski, D. A., Simons, J. R., Gu, J., Jamison, A. M., & Abroms, L. C. (2023). The efficacy of Facebook's vaccine misinformation policies and architecture during the COVID-19 pandemic. *Science Advances*, *9*(37), eadh2132. https://doi.org/10.1126/sciadv.adh2132
Bryant, A. G., Narasimhan, S., Bryant-Comstock, K., & Levi, E. E. (2014). Crisis pregnancy center websites: Information, misinformation and disinformation. *Contraception*, *90*(6), 601–605.
Byrne, P. A., Ma, T., Mann, R. E., & Elzohairy, Y. (2016). Evaluation of the general deterrence capacity of recently implemented (2009–2010) low and zero BAC requirements for drivers in Ontario. *Accident Analysis & Prevention*, *88*, 56–67.
Cai, M., Luo, H., Meng, X., Cui, Y., & Wang, W. (2023). Network distribution and sentiment interaction: Information diffusion mechanisms between social bots and human users on social media. *Information Processing & Management*, *60*(2), 103197.
Carey, J. M., Guess, A. M., Loewen, P. J., Merkley, E., Nyhan, B., Phillips, J. B., & Reifler, J. (2022). The ephemeral effects of fact-checks on COVID-19 misperceptions in the United States, Great Britain and Canada. *Nature Human Behaviour*, *6*(2), 236–243.
Cavallo, J. (2019). Findings From ASCO's Second National Cancer Opinion Survey. *The ASCO Post*.
Center for an Informed Public, Digital Forensic Research Lab, Graphika, & Stanford Internet Observatory. (2021). *The Long Fuse: Misinformation and the 2020 Election*. Stanford Digital Repository: Election Integrity Partnership. https://purl.stanford.edu/tr171zs0069
Chen, B., Shao, J., Liu, K., Cai, G., Jiang, Z., Huang, Y., Gu, H., & Jiang, J. (2018). Does eating chicken feet with pickled peppers cause avian influenza? Observational case study on Chinese social media during the avian influenza A (H7N9) outbreak. *JMIR Public Health and Surveillance*, *4*(1), e8198.
Chen, L., Wang, X., & Peng, T.-Q. (2018). Nature and diffusion of gynecologic cancer–related misinformation on social media: Analysis of tweets. *Journal of Medical Internet Research*, *20*(10), e11515.
CHEQ. (2019). *The economic cost of bad actors on the internet*. University of Baltimore.
Chipidza, W., & Yan, J. (Kevin). (2022). The effectiveness of flagging content belonging to prominent individuals: The case of Donald Trump on Twitter. *Journal of the Association* *for Information Science and Technology*, *73*(11), 1641–1658. https://doi.org/10.1002/asi.24705

Chu, Z., Gianvecchio, S., Wang, H., & Jajodia, S. (2012). Detecting automation of twitter accounts: Are you a human, bot, or cyborg? *IEEE Transactions on Dependable and Secure Computing*, *9*(6), 811–824.

Chua, A. Y., & Banerjee, S. (2018). Intentions to trust and share online health rumors: An experiment with medical professionals. *Computers in Human Behavior*, *87*, 1–9.

Chuai, Y., Tian, H., Pröllochs, N., & Lenzini, G. (2024). Did the Roll-Out of Community Notes Reduce Engagement With Misinformation on X/Twitter? *Proceedings of the ACM on Human-Computer Interaction*, *8*(CSCW2), 1–52. https://doi.org/10.1145/3686967

Cinelli, M., De Francisci Morales, G., Galeazzi, A., Quattrociocchi, W., & Starnini, M. (2021). The echo chamber effect on social media. *Proceedings of the National Academy of Sciences*, *118*(9), e2023301118. https://doi.org/10.1073/pnas.2023301118

Clarke, G. M., Conti, S., Wolters, A. T., & Steventon, A. (2019). Evaluating the impact of healthcare interventions using routine data. *Bmj*, *365*.

Clayton, K., Blair, S., Busam, J. A., Forstner, S., Glance, J., Green, G., Kawata, A., Kovvuri, A., Martin, J., & Morgan, E. (2020). Real solutions for fake news? Measuring the effectiveness of general warnings and fact-check tags in reducing belief in false stories on social media. *Political Behavior*, *42*, 1073–1095.

Coleman, K. (2021). *Introducing Birdwatch, a community-based approach to misinformation*. https://blog.x.com/en_us/topics/product/2021/introducing-birdwatch-a-community-based-approach-to-misinformation

Conger, K. (2020, October 9). Twitter Will Turn Off Some Features to Fight Election Misinformation. *The New York Times*. https://www.nytimes.com/2020/10/09/technology/twitter-election-ban-features.html

Crowell, C. (2017). *Our approach to bots and misinformation*. Twitter. https://blog.twitter.com/en_us/topics/company/2017/Our-Approach-Bots-Misinformation

Czerniak, K., Pillai, R., Parmar, A., Ramnath, K., Krocker, J., & Myneni, S. (2023). A scoping review of digital health interventions for combating COVID-19 misinformation and disinformation. *Journal of the American Medical Informatics Association*, *30*(4), 752–760.

D'arcy, J., & Herath, T. (2011). A review and analysis of deterrence theory in the IS security literature: Making sense of the disparate findings. *European Journal of Information Systems*, *20*(6), 643–658.

Dekens, N. (2023). *A Practical Guide for OSINT Investigators to Combat Disinformation and Fake Reviews Driven by AI (ChatGPT)*. Cheyenne, WY: Shadow Dragon.

Del Vicario, M., Bessi, A., Zollo, F., Petroni, F., Scala, A., Caldarelli, G., Stanley, H. E., & Quattrociocchi, W. (2016). The spreading of misinformation online. *Proceedings of the National Academy of Sciences*, *113*(3), 554–559. https://doi.org/10.1073/pnas.1517441113

DiFonzo, N., Robinson, N. M., Suls, J. M., & Rini, C. (2012). Rumors about cancer: Content, sources, coping, transmission, and belief. *Journal of Health Communication*, *17*(9), 1099–1115.

Dunn, A. G., Surian, D., Dalmazzo, J., Rezazadegan, D., Steffens, M., Dyda, A., Leask, J., Coiera, E., Dey, A., & Mandl, K. D. (2020). Limited role of bots in spreading vaccine-critical information among active twitter users in the United States: 2017–2019. *American Journal of Public Health*, *110*(S3), S319–S325.

Egli, A., Rosati, P., Lynn, T., & Sinclair, G. (2021). Bad Robot: A Preliminary Exploration of the Prevalence of Automated Software Programmes and Social Bots in the COVID-19# antivaxx Discourse on Twitter. *Proceedings of the The International Conference on Digital Society, Nice, France*, 18–22.

Facebook. (n.d.). *Fact-Checking Policies on Facebook, Instagram, and Threads*. Meta Business Help Center. Retrieved June 25, 2025, from https://www.facebook.com/business/help/315131736305613

Feiner, L. (2019, October 30). *Twitter bans political ads after Facebook refused to do so*. CNBC. https://www.cnbc.com/2019/10/30/twitter-bans-political-ads-after-facebook-refused-to-do-so.html

Fernández-Delgado, M., Cernadas, E., Barro, S., & Amorim, D. (2014). Do we need hundreds of classifiers to solve real world classification problems? *The Journal of Machine Learning Research*, *15*(1), 3133–3181.

Ford, C. L., Wallace, S. P., Newman, P. A., Lee, S.-J., & Cunningham, E. (2013). Belief in AIDS-related conspiracy theories and mistrust in the government: Relationship with HIV testing among at-risk older adults. *The Gerontologist*, *53*(6), 973–984.

Fried, I. (2019, October 17). *The changing face of disinformation: It's increasingly being spread by willing human actors*. Axios. https://www.axios.com/2019/10/17/misinformation-disinformation-human-actors

Gallwitz, F., & Kreil, M. (2022). Investigating the Validity of Botometer-Based Social Bot Studies. *Multidisciplinary International Symposium on Disinformation in Open Online Media*, 63–78.

Gefen, D., Ben-Assuli, O., Stehr, M., Rosen, B., & Denekamp, Y. (2019). Governmental intervention in Hospital Information Exchange (HIE) diffusion: A quasi-experimental ARIMA interrupted time series analysis of monthly HIE patient penetration rates. *European Journal of Information Systems*, *28*(6), 627–645. https://doi.org/10.1080/0960085X.2019.1666038

Georgiadou, E., Angelopoulos, S., & Drake, H. (2020). Big data analytics and international negotiations: Sentiment analysis of Brexit negotiating outcomes. *International Journal of Information Management*, *51*, 102048.

Gilani, Z., Farahbakhsh, R., Tyson, G., Wang, L., & Crowcroft, J. (2017). Of Bots and Humans (on Twitter). *Proceedings of the 2017 IEEE/ACM International Conference on Advances in Social Networks Analysis and Mining 2017*, 349–354. https://doi.org/10.1145/3110025.3110090

Gilbert, D. T., Tafarodi, R. W., & Malone, P. S. (1993). You can't not believe everything you read. *Journal of Personality and Social Psychology*, *65*(2), 221.

Google. (n.d.). *Misinformation policies (Youtube)*. Retrieved June 25, 2025, from https://support.google.com/youtube/answer/10834785?hl=en

Hagen, L., Neely, S., Keller, T. E., Scharf, R., & Vasquez, F. E. (2022). Rise of the Machines? Examining the Influence of Social Bots on a Political Discussion Network. *Social Science Computer Review*, *40*(2), 264–287. https://doi.org/10.1177/0894439320908190

Hamilton, J. D. (2020). *Time series analysis*. Princeton university press. https://books.google.com/books?hl=en&lr=&id=BeryDwAAQBAJ&oi=fnd&pg=PP1&ots=BhyO84cQlq&sig=A9Cd0Cq1IBOLHvYENPGGLTxq2ZE

Hanbury, A., Farley, K., Thompson, C., Wilson, P. M., Chambers, D., & Holmes, H. (2013). Immediate versus sustained effects: Interrupted time series analysis of a tailored intervention. *Implementation Science*, *8*, 1–9.

Harvey, D. (2019). *Helping you find reliable public health information on Twitter*. Twitter. https://blog.x.com/en_us/topics/company/2019/helping-you-find-reliable-public-health-information-on-twitter

Harvey, D., & Gasca, D. (2018). *Serving healthy conversation* [Twitter]. https://blog.x.com/en_us/topics/product/2018/Serving_Healthy_Conversation

Hassani, H., & Yeganegi, M. R. (2020). Selecting optimal lag order in Ljung–Box test. *Physica A: Statistical Mechanics and Its Applications*, *541*, 123700.

Hayawi, K., Saha, S., Masud, M. M., Mathew, S. S., & Kaosar, M. (2023). Social media bot detection with deep learning methods: A systematic review. *Neural Computing and Applications*, *35*(12), 8903–8918.

Hensel, P. G., & Kacprzak, A. (2021). Curbing cyberloafing: Studying general and specific deterrence effects with field evidence. *European Journal of Information Systems*, *30*(2), 219–235.

HHS. (2023). *National Health Observances | health.gov*. https://health.gov/news/category/national-health-observances

Higgins, G. E., Wilson, A. L., & Fell, B. D. (2005). An application of deterrence theory to software piracy. *Journal of Criminal Justice and Popular Culture*, *12*(3), 166–184.

Hill, J. A., Agewall, S., Baranchuk, A., Booz, G. W., Borer, J. S., Camici, P. G., Chen, P.-S., Dominiczak, A. F., Erol, Ç., & Grines, C. L. (2019). Medical misinformation: Vet the message! In *Journal of the American Heart Association* (Vol. 8, Issue 3, p. e011838). Am Heart Assoc.

Himelein-Wachowiak, M., Giorgi, S., Devoto, A., Rahman, M., Ungar, L., Schwartz, H. A., Epstein, D. H., Leggio, L., & Curtis, B. (2021). Bots and Misinformation Spread on Social Media: Implications for COVID-19. *Journal of Medical Internet Research*, *23*(5), e26933. https://doi.org/10.2196/26933

Ho, E. (2017). *An Update on Safety*. https://blog.x.com/en_us/topics/product/2017/an-update-on-safety

Horta Ribeiro, M., Cheng, J., & West, R. (2023). Automated Content Moderation Increases Adherence to Community Guidelines. *Proceedings of the ACM Web Conference 2023*, 2666–2676. https://doi.org/10.1145/3543507.3583275

Horta Ribeiro, M., Hosseinmardi, H., West, R., & Watts, D. J. (2023). Deplatforming did not decrease Parler users' activity on fringe social media. *PNAS Nexus*, *2*(3), pgad035.

Hu, K. (2023, February 2). ChatGPT sets record for fastest-growing user base—Analyst note. *Reuters*. https://www.reuters.com/technology/chatgpt-sets-record-fastest-growing-user-base-analyst-note-2023-02-01/

Hyndman, R. J., & Athanasopoulos, G. (2018). *Forecasting: Principles and practice*. OTexts.

Innes, H., & Innes, M. (2023). De-platforming disinformation: Conspiracy theories and their control. *Information, Communication & Society*, *26*(6), 1262–1280.

Instagram. (n.d.). *How Instagram addresses false info*. How Instagram Addresses False Info. Retrieved June 25, 2025, from https://help.instagram.com/2109682462659451?helpref=faq_content

Jamison, A. M., Broniatowski, D. A., & Quinn, S. C. (2019). Malicious actors on Twitter: A guide for public health researchers. *American Journal of Public Health*, *109*(5), 688–692.

Jhaver, S., Boylston, C., Yang, D., & Bruckman, A. (2021). Evaluating the effectiveness of deplatforming as a moderation strategy on Twitter. *Proceedings of the ACM on Human-Computer Interaction*, *5*(CSCW2), 1–30.

Johnson, N. F., Velásquez, N., Restrepo, N. J., Leahy, R., Gabriel, N., El Oud, S., Zheng, M., Manrique, P., Wuchty, S., & Lupu, Y. (2020). The online competition between pro-and anti-vaccination views. *Nature*, *582*(7811), 230–233.

Johnson, S. B., Park, H. S., Gross, C. P., & James, B. Y. (2018). Complementary medicine, refusal of conventional cancer therapy, and survival among patients with curable cancers. *JAMA Oncology*, *4*(10), 1375–1381.

Johnson, S. B., Park, H. S., Gross, C. P., & Yu, J. B. (2018). Use of alternative medicine for cancer and its impact on survival. *JNCI: Journal of the National Cancer Institute*, *110*(1), 121–124. https://doi.org/10.1093/jnci/djx145

Johnson, S. B., Parsons, M., Dorff, T., Moran, M. S., Ward, J. H., Cohen, S. A., Akerley, W., Bauman, J., Hubbard, J., & Spratt, D. E. (2022). Cancer misinformation and harmful information on Facebook and other social media: A brief report. *JNCI: Journal of the National Cancer Institute*, *114*(7), 1036–1039.

Jung, A.-K., Ross, B., & Stieglitz, S. (2020). Caution: Rumors ahead—A case study on the debunking of false information on Twitter. *Big Data & Society*, *7*(2), 205395172098012. https://doi.org/10.1177/2053951720980127

Kaiser, B., & Mayer, J. (2023). *It's the Algorithm: A large-scale comparative field study of misinformation interventions Emilie Flamme*. https://knightcolumbia.org/content/its-the-algorithm-a-large-scale-comparative-field-study-of-misinformation-interventions

Karami, A. (2025). A dual typology of social media interventions and deterrence mechanisms against misinformation. *Harvard Kennedy School Misinformation Review*. https://misinforeview.hks.harvard.edu/article/a-dual-typology-of-social-media-interventions-and-deterrence-mechanisms-against-misinformation/

Karami, A., Lundy, M., Webb, F., Turner-McGrievy, G., McKeever, B. W., & McKeever, R. (2021). Identifying and Analyzing Health-Related Themes in Disinformation Shared by Conservative and Liberal Russian Trolls on Twitter. *International Journal of Environmental Research and Public Health*, *18*(4), 2159. https://doi.org/10.3390/ijerph18042159

Karami, A., Qiao, Z., Zhang, X., Kharrazi, H., Bozorgi, P., & Bozorgi, A. (2024). Health Use Cases of AI Chatbots: Identification and Analysis of ChatGPT Prompts in Social Media Discourses. *Big Data and Cognitive Computing*, *8*(10), 130.

Karami, A., Zain, A., & Jamal, A. (2025). Unveiling the information mirage: A systematic literature review of health misinformation on social media. *Journal of Public Health*. https://doi.org/10.1007/s10389-025-02639-2

Katzowitz, J. (2018, August 6). *This grandmother tweets so much about Trump that Twitter thinks she's a bot*. The Daily Dot. https://www.dailydot.com/debug/twitter-bot-donald-trump-grandmother/

Kauk, J., Kreysa, H., & Schweinberger, S. R. (2025). Large-scale analysis of fact-checked stories on Twitter reveals graded effects of ambiguity and falsehood on information reappearance. *PNAS Nexus*, *4*(2), pgaf028.

Kemp, S. (2023). *Twitter Statistics and Trends*. DataReportal – Global Digital Insights. https://datareportal.com/essential-twitter-stats

Khaund, T., Kirdemir, B., Agarwal, N., Liu, H., & Morstatter, F. (2021). Social bots and their coordination during online campaigns: A survey. *IEEE Transactions on Computational Social Systems*, *9*(2), 530–545.

Kim, A., & Dennis, A. (2019). Says Who? The Effects of Presentation Format and Source Rating on Fake News in Social Media. *Management Information Systems Quarterly*, *43*(3), 1025–1039.

Kisa, A., & Kisa, S. (2025). Health conspiracy theories: A scoping review of drivers, impacts, and countermeasures. *International Journal for Equity in Health*, *24*(1), 93. https://doi.org/10.1186/s12939-025-02451-0

Koch, T. K., Frischlich, L., & Lermer, E. (2023). Effects of fact-checking warning labels and social endorsement cues on climate change fake news credibility and engagement on social media. *Journal of Applied Social Psychology*, *53*(6), 495–507. https://doi.org/10.1111/jasp.12959

Kornides, M. L., Badlis, S., Head, K. J., Putt, M., Cappella, J., & Gonzalez-Hernadez, G. (2023). Exploring content of misinformation about HPV vaccine on twitter. *Journal of Behavioral Medicine*, *46*(1–2), 239–252. https://doi.org/10.1007/s10865-022-00342-1

Laato, S., Islam, A. K. M. N., Islam, M. N., & Whelan, E. (2020). What drives unverified information sharing and cyberchondria during the COVID-19 pandemic? *European Journal of Information Systems*, *29*(3), 288–305. https://doi.org/10.1080/0960085X.2020.1770632

Lane, P. C., Clarke, D., & Hender, P. (2012). On developing robust models for favourability analysis: Model choice, feature sets and imbalanced data. *Decision Support Systems*, *53*(4), 712–718.

Lee, C.-B., & Teske Jr, R. H. (2015). Specific deterrence, community context, and drunk driving: An event history analysis. *International Journal of Offender Therapy and Comparative Criminology*, *59*(3), 230–258.

Levine, T. R. (2014). Truth-default theory (TDT) a theory of human deception and deception detection. *Journal of Language and Social Psychology*, *33*(4), 378–392.

Lewandowsky, S., Ecker, U. K. H., Seifert, C. M., Schwarz, N., & Cook, J. (2012). Misinformation and Its Correction: Continued Influence and Successful Debiasing. *Psychological Science in the Public Interest*, *13*(3), 106–131. https://doi.org/10.1177/1529100612451018

Li, S., Yang, J., & Zhao, K. (2023). *Are you in a Masquerade? Exploring the Behavior and Impact of Large Language Model Driven Social Bots in Online Social Networks* (No. arXiv:2307.10337). arXiv. http://arxiv.org/abs/2307.10337

Ljung, G. M., & Box, G. E. (1978). On a measure of lack of fit in time series models. *Biometrika*, *65*(2), 297–303.

Loeb, S., Sengupta, S., Butaney, M., Macaluso Jr, J. N., Czarniecki, S. W., Robbins, R., Braithwaite, R. S., Gao, L., Byrne, N., & Walter, D. (2019). Dissemination of misinformative and biased information about prostate cancer on YouTube. *European Urology*, *75*(4), 564–567.

Luceri, L., Cardoso, F., & Giordano, S. (2021). Down the bot hole: Actionable insights from a one-year analysis of bot activity on Twitter. *First Monday*. https://doi.org/10.5210/fm.v26i3.11441

Lyons, B. A., Montgomery, J. M., Guess, A. M., Nyhan, B., & Reifler, J. (2021). Overconfidence in news judgments is associated with false news susceptibility.

*Proceedings of the National Academy of Sciences*, *118*(23), e2019527118. https://doi.org/10.1073/pnas.2019527118
Lyons, K. (2021). *Twitter launches Birdwatch, a fact-checking program intended to fight misinformation—The Verge*. https://www.theverge.com/2021/1/25/22248903/twitter-birdwatch-fact-checking-misinformation
McCabe, S. D., Ferrari, D., Green, J., Lazer, D. M., & Esterling, K. M. (2024). Post-January 6th deplatforming reduced the reach of misinformation on Twitter. *Nature*, *630*(8015), 132–140.
McCleary, R., Hay, R. A., Meidinger, E. E., & McDowall, D. (1980). *Applied time series analysis for the social sciences*. Sage Publications Beverly Hills, CA.
McHugh, M. L. (2012). Interrater reliability: The kappa statistic. *Biochemia Medica: Biochemia Medica*, *22*(3), 276–282.
Mehta, S. (2023). Towards Informed Interventions to Limit the Effects of Misleading Information on Social Networks [Ph.D., New York University]. In *ProQuest Dissertations and Theses* (2884000734). ProQuest Dissertations & Theses Global. https://uab.idm.oclc.org/login?url=https://www.proquest.com/dissertations-theses/towards-informed-interventions-limit-effects/docview/2884000734/se-2?accountid=8240
Mena, P. (2020). Cleaning Up Social Media: The Effect of Warning Labels on Likelihood of Sharing False News on Facebook. *Policy & Internet*, *12*(2), 165–183. https://doi.org/10.1002/poi3.214
Meta. (2025a). *How fact-checking works*. How Fact-Checking Works. https://transparency.meta.com/features/how-fact-checking-works
Meta. (2025b). *Misinformation*. https://transparency.meta.com/policies/community-standards/misinformation
Moravec, P., Minas, R., & Dennis, A. R. (2019). Fake news on social media: People believe what they want to believe when it makes no sense at all. *MIS Quarterly*, *43*(4).
Mudambi, M., Clark, J., Rhue, L., & Viswanathan, S. (2025). Fighting Misinformation on Social Media: An Empirical Investigation of the Impact of Prominence Reduction Policies. *Production and Operations Management*, *34*(6), 1409–1425. https://doi.org/10.1177/10591478241283839
Muhammed T, S., & Mathew, S. K. (2022). The disaster of misinformation: A review of research in social media. *International Journal of Data Science and Analytics*, *13*(4), 271–285. https://doi.org/10.1007/s41060-022-00311-6
Nagin, D. S. (1998). Criminal deterrence research at the outset of the twenty-first century. *Crime and Justice*, *23*, 1–42.
Ng, K. C., Tang, J., & Lee, D. (2021). The effect of platform intervention policies on fake news dissemination and survival: An empirical examination. *Journal of Management Information Systems*, *38*(4), 898–930.
Nye, J. S. (2005). *Soft power: The means to success in world politics*. Public affairs. https://www.wcfia.harvard.edu/publications/soft-power-means-success-world-politics
Okuhara, T., Ishikawa, H., Okada, M., Kato, M., & Kiuchi, T. (2017). Assertions of Japanese websites for and against cancer screening: A text mining analysis. *Asian Pacific Journal of Cancer Prevention: APJCP*, *18*(4), 1069.

Papakyriakopoulos, O., Serrano, J. C. M., & Hegelich, S. (2020). The spread of COVID-19 conspiracy theories on social media and the effect of content moderation. *Harvard Kennedy School Misinformation Review*, *1*(3).

Pennycook, G., & Rand, D. G. (2019). Fighting misinformation on social media using crowdsourced judgments of news source quality. *Proceedings of the National Academy of Sciences*, *116*(7), 2521–2526.

Persily, N. (2017). Can democracy survive the Internet? *J. Democracy*, *28*, 63.

Pham, B. T., Bui, D. T., & Prakash, I. (2017). Landslide susceptibility assessment using bagging ensemble based alternating decision trees, logistic regression and J48 decision trees methods: A comparative study. *Geotechnical and Geological Engineering*, *35*(6), 2597–2611.

Porter, E., & Wood, T. J. (2021). The global effectiveness of fact-checking: Evidence from simultaneous experiments in Argentina, Nigeria, South Africa, and the United Kingdom. *Proceedings of the National Academy of Sciences*, *118*(37), e2104235118. https://doi.org/10.1073/pnas.2104235118

Radivojevic, K., Clark, N., & Brenner, P. (2024). LLMs Among Us: Generative AI Participating in Digital Discourse. *Proceedings of the AAAI Symposium Series*, *3*(1), 209–218. https://ojs.aaai.org/index.php/AAAI-SS/article/view/31202

Rahwan, I., Cebrian, M., Obradovich, N., Bongard, J., Bonnefon, J.-F., Breazeal, C., Crandall, J. W., Christakis, N. A., Couzin, I. D., & Jackson, M. O. (2019). Machine behaviour. *Nature*, *568*(7753), 477–486.

Rockwell, P. (2022). *An update on how we keep members safe on LinkedIn*. https://www.linkedin.com/blog/member/trust-and-safety/an-update-on-how-we-keep-members-safe-on-linkedin

Roozenbeek, J., van der Linden, S., & Nygren, T. (2020). Prebunking interventions based on "inoculation" theory can reduce susceptibility to misinformation across cultures. *Harvard Kennedy School Misinformation Review*, *1*(2). https://misinforeview.hks.harvard.edu/article/global-vaccination-badnews/

Ross, B., Jung, A., Heisel, J., & Stieglitz, S. (2018). *Fake News on Social Media: The (In) Effectiveness of Warning Messages*.

Ross, B., Pilz, L., Cabrera, B., Brachten, F., Neubaum, G., & Stieglitz, S. (2019). Are social bots a real threat? An agent-based model of the spiral of silence to analyse the impact of manipulative actors in social networks. *European Journal of Information Systems*, *28*(4), 394–412. https://doi.org/10.1080/0960085X.2018.1560920

Roth, Y., & Harvey, D. (2018). *How Twitter is fighting spam and malicious automation*. Twitter. https://blog.x.com/en_us/topics/company/2018/how-twitter-is-fighting-spam-and-malicious-automation

Roth, Y., & Pickles, N. (2020). *Updating our approach to misleading information*. Twitter. https://blog.x.com/en_us/topics/product/2020/updating-our-approach-to-misleading-information

Saan, A. (2024, July 16). Twitter API Changes: Navigating the End of Free Access | Your 2024 Guide. *Medium*. https://medium.com/@asaan/twitter-api-changes-navigating-the-end-of-free-access-your-2024-guide-b9f9cf47ea79

Said, S. E., & Dickey, D. A. (1984). Testing for unit roots in autoregressive-moving average models of unknown order. *Biometrika*, *71*(3), 599–607.

Salge, C. A. de L., Karahanna, E., & Thatcher, J. B. (2022). Algorithmic Processes of Social Alertness and Social Transmission: How Bots Disseminate Information on Twitter. *Mis Quarterly*, *46*(1). https://search.ebscohost.com/login.aspx?direct=true&profile=ehost&scope=site&authtype=crawler&jrnl=02767783&AN=155825474&h=swdM0qEOXSV35kzFXrLlqGy3GTDcufv76TbcS1vS%2BMy%2FsHiB1HvlLwk2jX8x9AHOrW36fs5FsN%2FZb%2BnaI4IiAA%3D%3D&crl=c

Sanderson, Z., Brown, M. A., Bonneau, R., Nagler, J., & Tucker, J. A. (2021). Twitter flagged Donald Trump's tweets with election misinformation: They continued to spread both on and off the platform. *Harvard Kennedy School Misinformation Review*.

Scerbo, M. (2007). Adaptive automation. *Neuroergonomics: The Brain at Work*, *239252*. https://books.google.com/books?hl=en&lr=&id=9ERRDAAAQBAJ&oi=fnd&pg=PA239&ots=bx7fFoetg6&sig=Q_oWNuUCRZEjOAtcMVhd-_iyRYY

Schaffer, A. L., Dobbins, T. A., & Pearson, S.-A. (2021). Interrupted time series analysis using autoregressive integrated moving average (ARIMA) models: A guide for evaluating large-scale health interventions. *BMC Medical Research Methodology*, *21*(1), 1–12.

Schuetz, S. W., Sykes, T. A., & Venkatesh, V. (2021). Combating COVID-19 fake news on social media through fact checking: Antecedents and consequences. *European Journal of Information Systems*, *30*(4), 376–388. https://doi.org/10.1080/0960085X.2021.1895682

Seering, J., Kraut, R., & Dabbish, L. (2017). Shaping pro and anti-social behavior on twitch through moderation and example-setting. *Proceedings of the 2017 ACM Conference on Computer Supported Cooperative Work and Social Computing*, 111–125.

Sehat, C. M., Li, R., Nie, P., Prabhakar, T., & Zhang, A. X. (2024). Misinformation as a Harm: Structured Approaches for Fact-Checking Prioritization. *Proceedings of the ACM on Human-Computer Interaction*, *8*(CSCW1), 1–36. https://doi.org/10.1145/3641010

Shao, C., Ciampaglia, G. L., Varol, O., Flammini, A., & Menczer, F. (2017). The spread of fake news by social bots. *arXiv Preprint arXiv:1707.07592*, *96*, 104.

Shao, C., Ciampaglia, G. L., Varol, O., Yang, K.-C., Flammini, A., & Menczer, F. (2018). The spread of low-credibility content by social bots. *Nature Communications*, *9*(1), 1–9.

Sharma, P. R., Spearing, E. R., Wade, K. A., & Jobson, L. (2024). Distress reactions and susceptibility to misinformation for an analogue trauma event. *Cognitive Research: Principles and Implications*, *9*(1), 53. https://doi.org/10.1186/s41235-024-00582-6

Shewale, R. (2023). *Twitter Statistics In 2023*. https://www.demandsage.com/twitter-statistics/#:~:text=Let%20us%20take%20a%20closer,528.3%20million%20monthly%20active%20users

Shin, D., Kee, K. F., & Shin, E. Y. (2023). The Nudging Effect of Accuracy Alerts for Combating the Diffusion of Misinformation: Algorithmic News Sources, Trust in Algorithms, and Users' Discernment of Fake News. *Journal of Broadcasting & Electronic Media*, *67*(2), 141–160. https://doi.org/10.1080/08838151.2023.2175830

Shu, K., Wang, S., Lee, D., & Liu, H. (2020). *Mining Disinformation and Fake News: Concepts, Methods, and Recent Advancements*. Springer.

Shumway, R. H., Stoffer, D. S., & Stoffer, D. S. (2000). *Time series analysis and its applications* (Vol. 3). Springer.

Siebert, J., & Siebert, J. U. (2023). Effective mitigation of the belief perseverance bias after the retraction of misinformation: Awareness training and counter-speech. *Plos One*, *18*(3), e0282202.

Slaughter, I., Peytavin, A., Ugander, J., & Saveski, M. (2025). *Community Notes Moderate Engagement With and Diffusion of False Information Online* (No. arXiv:2502.13322). arXiv. https://doi.org/10.48550/arXiv.2502.13322

Son, J., Lee, H. K., Jin, S., & Lee, J. (2019). Content features of tweets for effective communication during disasters: A media synchronicity theory perspective. *International Journal of Information Management*, *45*, 56–68.

Srikanth, A. (2021, March 24). 12 prominent people opposed to vaccines are responsible for two-thirds of anti-vaccine content online: Report [Text]. *The Hill*. https://thehill.com/changing-america/well-being/prevention-cures/544712-twelve-anti-vaxxers-are-responsible-for-two/

Stafford, M. C., & Warr, M. (1993). A reconceptualization of general and specific deterrence. *Journal of Research in Crime and Delinquency*, *30*(2), 123–135.

Tandoc Jr, E. C. (2019). The facts of fake news: A research review. *Sociology Compass*, *13*(9), e12724.

The Lancet Infectious Diseases. (2020). The COVID-19 infodemic. *The Lancet. Infectious Diseases*, *20*(8), 875. https://doi.org/10.1016/S1473-3099(20)30565-X

Théro, H., & Vincent, E. M. (2022). Investigating Facebook's interventions against accounts that repeatedly share misinformation. *Information Processing & Management*, *59*(2), 102804.

Thorson, E. (2016). Belief echoes: The persistent effects of corrected misinformation. *Political Communication*, *33*(3), 460–480.

Tomlinson, K. D. (2016). An examination of deterrence theory: Where do we stand. *Fed. Probation*, *80*, 33.

Tran, T., Valecha, R., & Rao, H. R. (2023). Machine and human roles for mitigation of misinformation harms during crises: An activity theory conceptualization and validation. *International Journal of Information Management*, *70*, 102627.

Truong, B. T., Kim, S., Nogara, G., Verdolotti, E., Sahneh, E. S., Saurwein, F., Just, N., Luceri, L., Giordano, S., & Menczer, F. (2025). Delayed takedown of illegal content on social media makes moderation ineffective. *arXiv Preprint arXiv:2502.08841*. https://arxiv.org/abs/2502.08841

Tsay, R. S. (2005). *Analysis of financial time series*. John wiley & sons.

Twitter. (2017). *Update: Russian interference in the 2016 US presidential election*. Twitter. https://blog.x.com/en_us/topics/company/2017/Update-Russian-Interference-in-2016--Election-Bots-and-Misinformation

Twitter. (2020a). *Expanding our policies to further protect the civic conversation*. https://blog.x.com/en_us/topics/company/2020/civic-integrity-policy-update

Twitter. (2020b). *Updates to our work on COVID-19 vaccine misinformation*. https://blog.x.com/en_us/topics/company/2021/updates-to-our-work-on-covid-19-vaccine-misinformation

Twitter. (2021a). *Coronavirus: Staying safe and informed on Twitter*. https://blog.x.com/en_us/topics/company/2020/covid-19

Twitter. (2021b). *Platform Manipulation*. https://transparency.twitter.com/en/reports/platform-manipulation.html

Twitter. (2023). *Synthetic and manipulated media policy*. https://help.twitter.com/en/rules-and-policies/manipulated-media

Varol, O., Ferrara, E., Davis, C., Menczer, F., & Flammini, A. (2017). Online human-bot interactions: Detection, estimation, and characterization. *Proceedings of the International AAAI Conference on Web and Social Media*, *11*(1), 280–289.

Varzgani, H., Kordzadeh, N., & Lee, K. (2023). Toward Designing Effective Warning Labels for Health Misinformation on Social Media. *Proceedings of the 56th Hawaii International Conference on System Sciences*. HICSS 2023. https://scholarspace.manoa.hawaii.edu/items/30ebd144-6790-4abc-bb0e-c1896009b0c9

Vasconcelos Silva, C., Jayasinghe, D., & Janda, M. (2020). What can Twitter tell us about skin cancer Communication and prevention on social media? *Dermatology*, *236*(2), 81–89.

Vincent, E. M., Théro, H., & Shabayek, S. (2022). Measuring the effect of Facebook's downranking interventions against groups and websites that repeatedly share misinformation. *Harvard Kennedy School Misinformation Review*.

Vosoughi, S., Roy, D., & Aral, S. (2018). The spread of true and false news online. *Science*, *359*(6380), 1146–1151.

Wang, P., Angarita, R., & Renna, I. (2018). Is this the era of misinformation yet: Combining social bots and fake news to deceive the masses. *Companion Proceedings of the The Web Conference 2018*, 1557–1561.

Wang, Y., McKee, M., Torbica, A., & Stuckler, D. (2019). Systematic literature review on the spread of health-related misinformation on social media. *Social Science & Medicine*, *240*, 112552. https://doi.org/10.1016/j.socscimed.2019.112552

Warner, E. L., Basen-Engquist, K. M., Badger, T. A., Crane, T. E., & Raber-Ramsey, M. (2022). The Online Cancer Nutrition Misinformation: A framework of behavior change based on exposure to cancer nutrition misinformation. *Cancer*. https://doi.org/10.1002/cncr.34218

Williams, K. R., & Hawkins, R. (1986). Perceptual research on general deterrence: A critical review. *LAw & Soc'Y REv.*, *20*, 545.

Wilner, T., & Holton, A. (2020). Breast cancer prevention and treatment: Misinformation on Pinterest, 2018. *American Journal of Public Health*, *110*(S3), S300–S304. https://doi.org/10.2105/AJPH.2020.305812

Wischnewski, M., Ngo, T., Bernemann, R., Jansen, M., & Krämer, N. (2024). "I agree with you, bot!" How users (dis)engage with social bots on Twitter. *New Media & Society*, *26*(3), 1505–1526. https://doi.org/10.1177/14614448211072307

Wu, L., Morstatter, F., Carley, K. M., & Liu, H. (2019). Misinformation in Social Media: Definition, Manipulation, and Detection. *ACM SIGKDD Explorations Newsletter*, *21*(2), 80–90. https://doi.org/10.1145/3373464.3373475

Wu, X., Kumar, V., Quinlan, J. R., Ghosh, J., Yang, Q., Motoda, H., McLachlan, G. J., Ng, A., Liu, B., & Philip, S. Y. (2008). Top 10 algorithms in data mining. *Knowledge and Information Systems*, *14*(1), 1–37.

X. (2023). *Not A Bot*. https://help.x.com/en/using-x/not-a-bot

X. (2025). *Authenticity*. Authenticity. https://help.x.com/en/rules-and-policies/authenticity

Xu, A. J., Taylor, J., Gao, T., Mihalcea, R., Perez-Rosas, V., & Loeb, S. (2021). TikTok and prostate cancer: Misinformation and quality of information using validated questionnaires. *BJU International*, *128*(4), 435–437.

Xu, Z., & Guo, H. (2018). Using text mining to compare online pro-and anti-vaccine headlines: Word usage, sentiments, and online popularity. *Communication Studies*, *69*(1), 103–122.

Yang, K.-C., & Menczer, F. (2024). Anatomy of an AI-powered malicious social botnet. *Journal of Quantitative Description: Digital Media*, *4*. https://doi.org/10.51685/jqd.2024.icwsm.7

Yang, K.-C., Varol, O., Davis, C. A., Ferrara, E., Flammini, A., & Menczer, F. (2019). Arming the public with artificial intelligence to counter social bots. *Human Behavior and Emerging Technologies*, *1*(1), 48–61.

Young, V. A. (2020). *Nearly Half of the Twitter Accounts Discussing "Reopening America" May Be Bots*. http://www.cmu.edu/news/stories/archives/2020/may/twitter-bot-campaign.html

Zannettou, S. (2021). " I Won the Election!": An Empirical Analysis of Soft Moderation Interventions on Twitter. *Proceedings of the International AAAI Conference on Web and Social Media*, *15*, 865–876.

Zhai, W., Yu, H., & Song, C. Y. (2024). Disaster Misinformation and Its Corrections on Social Media: Spatiotemporal Proximity, Social Network, and Sentiment Contagion. *Annals of the American Association of Geographers*, *114*(2), 408–435. https://doi.org/10.1080/24694452.2023.2271549

Zhang, C., Liu, C., Zhang, X., & Almpanidis, G. (2017). An up-to-date comparison of state-of-the-art classification algorithms. *Expert Systems with Applications*, *82*, 128–150.

Zhang, J., Zhang, R., Zhang, Y., & Yan, G. (2016). The rise of social botnets: Attacks and countermeasures. *IEEE Transactions on Dependable and Secure Computing*, *15*(6), 1068–1082.

Zhang, Y., Song, W., Shao, J., Abbas, M., Zhang, J., Koura, Y. H., & Su, Y. (2023). Social bots' role in the COVID-19 pandemic discussion on Twitter. *International Journal of Environmental Research and Public Health*, *20*(4), 3284.

Zhong, W., Broniatowski, D., Dredze, M., & Abroms, L. (2023). *Evaluating Twitter's COVID-19 Vaccine Misinformation Removal Policy*. https://osf.io/preprints/socarxiv/cxg6y/

## Appendices

### Appendix A Statistical Analysis

We used a seasonal ARIMA model expressed as $(p, d, q) \times (P, D, Q)_S$. Primary ARIMA model consists of three components: $p$ representing the order of autoregressive (AR) terms, $d$ indicating the degree of non-seasonal differencing, and $q$ denoting moving average (MA) terms. A seasonal ARIMA model extends the basic ARIMA model to handle seasonal data patterns. The model integrates both non-seasonal and seasonal elements to provide a comprehensive approach for analyzing and forecasting time series data with evident seasonality. The notation $(p, d, q)$ captures the non-seasonal components, while $(P, D, Q)_S$ represents the seasonal components. In this model, P represents the order of the seasonal autoregressive (AR) terms, which capture the relationship between an observation and its seasonal lagged observations. Similar to the primary components, $P$ represents the order of seasonal autoregressive or AR, D shows the degree of seasonal differencing, Q denotes seasonal moving average or MA, and S represents the number of observation per year (Schaffer et al., 2021). The seasonal part of the model, therefore, effectively captures and adjusts for recurring patterns and variations that occur at regular intervals, ensuring more accurate and reliable forecasts for data with strong seasonal influences (Brockwell & Davis, 2002).

Before determining the most suitable models, it is necessary to transform the time series into a stationary state, indicating constancy in mean, variance, and autocorrelation function (ACF) (Schaffer et al., 2021). However, the results of the augmented Dicky-Fuller test (Said & Dickey, 1984) indicated that the analyzed time series was non-stationary. Non-stationarity in time series data can stem from two primary sources. The first is varying variance over time, known as heteroscedasticity, which can often be mitigated by applying a logarithmic transformation. The second source is an upward or downward trend, which can often be eliminated by taking the first difference between consecutive observations (Hyndman & Athanasopoulos, 2018). Therefore, we used log-transformation and differentiation. We used the R squared, root mean square error (RMSE), mean absolute error (MAE), and the Ljung-Box test (Ljung & Box, 1978) to assess ARIMA models. The test helps determine whether the observed autocorrelation in the data differs significantly from what would be expected under the assumption of no autocorrelation. The test checks if errors in the data randomly bounce around, like coin flips, or if they're connected across time, hinting at hidden patterns. No connection suggests randomness, while connections mean there's more to the data's story. Not rejecting the null hypothesis (p-vlaue > 0.05) in the Ljung-Box test implies that the model has successfully accounted for the remaining autocorrelation in the data, indicating a good fit (the observed connection between errors is random). To define the degree of freedom (df) for the Ljung-Box test, we used different values for df, including 5, ln(T) (Tsay, 2005), 20 (Shumway et al., 2000), min(10, T/5) (Hyndman & Athanasopoulos, 2018), and 3 (Hassani & Yeganegi, 2020), where T represents the number of observations.

To evaluate the intervention's effect, we examined the coefficients associated with the intervention parameters and tested their statistical significance. The magnitude of the effect was assessed by comparing the actual observations with the predicted data pattern in the absence of the intervention (Box & Jenkins, 1976). In this study, the coefficients of the intervention parameters indicated the change in the average monthly misinformation posted on social media following the intervention start point. Assuming that the parameter for the intervention deterrence effect is $\beta$, we measured the effect percentage using $(10^{\beta} - 1) \times 100$.

**Appendix B Post Hoc Analyses**

We extended our analyses to better compare this study with previous research on the deterrence effect of social media interventions in curbing the spread of misinformation without differentiating between bots and humans. Table B1 presents a detailed summary of the ARIMA model used to examine the effects of interventions on both bot and human activities related to the spread of health misinformation. Our result indicates that the overall volume of misinformation generated by both bots and humans dropped significantly (coefficient = -1.094, p-value = 0.001) after starting the removal strategy. This suggests that the removal intervention successfully reduced the spread of health misinformation over time across all users. However, the impacts of the reduction and informing strategies were not significant. This shows that the two strategies were not successful in reducing health misinformation. The Ljung-Box statistics at various degrees of freedom revealed high p-values, confirming that there was no significant autocorrelation in the residuals.

Table B1 Summary of ARIMA intervention model estimation for bot and human activities.

| **Model Estimation: Activities of All Users (Bot + Human)** | | | | |
|---|---|---|---|---|
| Variable | Estimate | SE | t | *p-value* |
| Removal | -1.094 | 0.315 | -3.475 | 0.001* |
| Reduction | -0.535 | 0.296 | -1.810 | 0.070 |
| Informing | -0.487 | 0.328 | -1.484 | 0.138 |
| **Model Assessment** | | | | |
| Stationary $R^2$ | 0.295 | | | |
| RMSE | 0.406 | | | |
| MAE | 0.296 | | | |
| Ljung-Box statistics | X-squared = 0.17434, df = 3, p-value = 0.9816 | | | |
| | X-squared = 0.97197, df = 4, p-value = 0.914 | | | |
| | X-squared = 1.7644, df = 5, p-value = 0.8807 | | | |
| | X-squared = 8.9404, df = 10, p-value = 0.5378 | | | |
| | X-squared = 9.9396, df = 12, p-value = 0.6213 | | | |
| | X-squared = 17.944, df = 20, p-value = 0.5911 | | | |

A thorough overview of the ARIMA model used to evaluate the effect of interventions on the quantity of bot and human users spreading health misinformation about health is provided in Table B2. The findings show that there was no statistically significant change in the number of users after the intervention strategies were put into place. This result implies that over time, the interventions did not successfully lower users involved in disseminating false information. Furthermore, the absence of significant autocorrelation in the model residuals was confirmed by the high p-values obtained from the Ljung-Box statistics across multiple lags.

Table B.2 Summary of ARIMA intervention model estimation for bot and human users.

| **Model Estimation: Activities of All Users (Bot + Human)** | | | | |
|---|---|---|---|---|
| Variable | Estimate | SE | t | *p-value* |
| Removal | 0.248 | 0.269 | 0.921 | 0.357 |
| Reduction | -0.003 | 0.261 | -0.011 | 0.991 |
| Informing | -0.304 | 0.279 | -1.092 | 0.275 |
| **Model Assessment** | | | | |
| Stationary $R^2$ | 0.373 | | | |

| | |
|---|---|
| RMSE | 0.379 |
| MAE | 0.290 |
| Ljung-Box statistics | X-squared = 0.75765, df = 3, p-value = 0.8596<br>X-squared = 0.91588, df = 4, p-value = 0.9223<br>X-squared = 1.3997, df = 5, p-value = 0.9243<br>X-squared = 2.5131, df = 10, p-value = 0.9907<br>X-squared = 2.6821, df = 12, p-value = 0.9974<br>X-squared = 9.0838, df = 20, p-value = 0.9819 |

This post hoc analysis, which examined both human and bot users and their activities, indicated that the removal strategy reduced the overall number of users but did not have a significant impact on activity levels. While the results for the removal strategy were mixed, the reduction and informing strategies also did not show a lasting deterrent effect on the number of users or their activities. However, the disaggregated results reveal a more nuanced picture: both the removal and reduction strategies effectively reduced bot activities (Table 5), and all three interventions decreased the number of bot users (Table 4). Although the aggregated results might suggest that the reduction strategy could be effective for human users, the disaggregated findings show that none of the interventions had a significant impact on the presence or activity levels of human users (Tables 6–7).